\documentclass[reprint,twocolumn,superscriptaddress,showpacs,preprintnumbers,nofootinbib,aps,prd]{revtex4-1}
\usepackage{graphicx,color}
\usepackage{graphicx}
\usepackage{amsmath,amssymb,amsfonts}
\usepackage{stackrel}
\usepackage{enumitem}
\usepackage[english]{babel}
\usepackage{verbatim}
\usepackage{comment}
\usepackage{ulem}
\usepackage{xcolor}
\usepackage{subcaption}
\usepackage{dcolumn}
\usepackage{physics}
\usepackage{verbatim}
\usepackage{float}
\usepackage{color}
\usepackage{multirow}
\usepackage{graphicx}
\usepackage{epstopdf}
\usepackage{epsfig}
\usepackage{slashed}
\usepackage{physics}
\usepackage{multirow}
\usepackage{makecell}

\definecolor{darkred}{rgb}{.7,.1,.1}
\usepackage{hyperref}
\hypersetup{
    pdfnewwindow=true,  
    colorlinks=true,     
    linkcolor=darkred,     
    citecolor=blue,      
    filecolor=blue,      
    urlcolor=blue        
}
 
\graphicspath{
  {./}
  {figures/}
}

\definecolor{dark-green}{rgb}{0.1,0.7,0.3}

\newcommand{\tf}{\texorpdfstring}
\newcommand{\gev}{~\text{GeV}}
\newcommand{\tev}{~\text{TeV}}

\newcommand{\fb}{~\text{fb}}

\newcommand{\fbi}{~\text{fb}^{-1}}
\newcommand{\abi}{~\text{ab}^{-1}}

\newcommand{\onbb}{0\nu\beta\beta}

\def\nn{\nonumber}

\newcommand{\SUNYATSET}{\affiliation{School of Physics and Astronomy, Sun Yat-sen University, Zhuhai 519082, P.R. China.}}
\newcommand{\CAS}{\affiliation{CAS Key Laboratory of Theoretical Physics, Institute of Theoretical Physics, Chinese Academy of Sciences,    \\ Beijing 100190, P. R. China}}
\newcommand{\ZKD}{\affiliation{School of Physical Sciences, University of Chinese Academy of Sciences,   Beijing 100049, P.R. China}}
\newcommand{\HEP}{\affiliation{Center for High Energy Physics, Peking University, Beijing 100871, China}}
\newcommand{\HZGYY}{\affiliation{School of Fundamental Physics and Mathematical Sciences, Hangzhou Institute for Advanced Study, UCAS, Hangzhou 310024, China}}
\newcommand{\ICFT}{\affiliation{International Centre for Theoretical Physics Asia-Pacific, Beijing/Hangzhou, China}}

\begin{document}
\title{Complementary LHC searches for UV resonances of the $0\nu\beta\beta$ decay operators}
\author{Gang Li}
\email{ligang65@mail.sysu.edu.cn}
\SUNYATSET
\author{Jiang-Hao Yu}
\email{jhyu@itp.ac.cn}
\CAS
\ZKD
\HEP
\HZGYY
\ICFT
\author{Xiang Zhao}
\email{zhaox88@mail2.sysu.edu.cn}
\SUNYATSET

\begin{abstract}
\vspace*{0.5cm}

We investigate the quark-level effective operators related to the neutrinoless double beta $(0\nu\beta\beta)$ decay process, and their ultraviolet completions relevant to chiral enhancement effects at the hadronic level. We have classified several kinds of leptoquark models, matching to different standard model effective operators. Assuming weakly-coupled new physics, we find the ongoing $0\nu\beta\beta$-decay experiments are sensitive to new physics scale at around $2\sim 4$ TeV, which is in the reach of LHC searches. We discuss the discovery potential of such resonances in the same-sign dilepton channels at the LHC. Therefore, the direct LHC searches and indirect $0\nu\beta\beta$-decay searches are complementary to each other in testing the UV completions of the effective operators for $0\nu\beta\beta$ decay.

\end{abstract}

\pacs{}
\maketitle

\section{Introduction}
\label{sec:intro}

The origin of the tiny neutrino masses remains mysterious in particle physics. To understand this mystery, it is crucial to know the nature of neutrinos, whether the neutrinos are Majorana or Dirac fermions. 
Neutrinoless double beta decay $(0\nu\beta\beta)$ in nuclei provides the most sensitive way to assess the Majorana nature of neutrinos~\cite{Furry:1939qr}. It has been shown~\cite{Schechter:1981bd} that if the $\onbb$ decay process is observed, neutrinos must have Majorana masses. 

There have been lots of studies on $\onbb$ decay, from the search for the signal in the standard mechanism with the exchange of three light Majorana neutrinos to the various interpretations in terms of new physics beyond the standard model (BSM), e.g. see Refs.~\cite{Rodejohann:2011mu,Agostini:2022zub} for review. 
In ultraviolet (UV) theories such as the left-right symmetric model (LRSM)~\cite{Mohapatra:1979ia,Mohapatra:1980yp}, the origin of neutrino masses is explained while promising signals in $\onbb$ decay and at colliders are predicted~\cite{Tello:2010am,Cirigliano:2018yza,Li:2020xlh,BhupalDev:2013ntw}. Besides the Majorana masses of neutrinos, $\onbb$ decay can also arise from other lepton-number-violating (LNV) sources, which may give negligible contributions to the neutrino masses.

Given the hierarchy between the new physics scale and the nuclei scale, the effective field theory (EFT) framework has been utilized~\cite{Prezeau:2003xn,Cirigliano:2017djv,Cirigliano:2018yza,Scholer:2023bnn} to systematically parametrize various new physics effects in the low-energy hadronic and nuclei scales, matching theories to experiments with unprecedented sensitivities.

Above the electroweak (EW) scale, the BSM physics can be systematically described by a series of higher dimensional operators in the standard model effective field theory (SMEFT), which respects the $SU(3)_C\times SU(2)_L \times U(1)_Y$ gauge symmetry. Effective operators with odd dimensions can give rise to the lepton number violation by two units, $\Delta L =2$. There is only one dim-5 (dimension 5) operator, that is the Weinberg operator~\cite{Weinberg:1979sa}, while the numbers of effective operators at higher dimensions blow up. The complete bases of dim-7~\cite{Lehman:2014jma,Liao:2016hru} and dim-9 operators~\cite{Li:2020xlh,Liao:2020jmn} in the SMEFT have been obtained in recent years. Below the EW scale, the SM heavy fields including Higgs boson, top quark, $W$ and $Z$ bosons are integrated out leading to the low-energy EFT (LEFT) with the unbroken $SU(3)_C\times U(1)_{em }$ gauge symmetry. The complete bases of effective operators in the LEFT up to dim-9 have been achieved~\cite{Jenkins:2017jig,Liao:2020zyx,Li:2020tsi}.

Below the hadronic scale, the quark-lepton operators should be matched to the chiral Lagrangian at the hadronic level, and these hadronic operators are organized by chiral power counting. If the $\onbb$-decay process originates from the exchange of BSM fields, the LNV interactions at low energies are described by the dim-9 quark-lepton operators in the LEFT. Among them, the operators $O_4 \bar{e}_L e_L^c$ and $O_4 \bar{e}_R e_R^c$  (cf. Eq.~\eqref{eq:O4-LEFT}) give rise to the long-range pion exchange at the hadronic scale with chiral enhancement~\cite{Prezeau:2003xn,Graesser:2016bpz}. The impact on the half-life of $\onbb$ decay has been studied in detail in Refs.~\cite{Prezeau:2003xn,Cirigliano:2018yza}.

Assuming weakly-coupled new physics, the future tonne-scale $0\nu\beta\beta$ experiments are typically sensitive to the new physics scale $\Lambda$ around $1\sim 2\tev$, which is in the reach of the searches at the Large Hadron Collider (LHC)~\cite{Helo:2013ika,Graesser:2022nkv}. However, the chiral enhancement effect would greatly enhance the $\onbb$-decay rate, and thus push the new physics scale $\Lambda$ to $2\sim 4\tev$.
This has been investigated in the LRSM~\cite{Li:2020flq} due to the large mixing between the left- and right-handed $W$ bosons, which renders the effective operators $O_4 \bar{e}_L e_L^c$ and $O_4 \bar{e}_R e_R^c$ after integrating out heavy particles below the EW scale. However, in this scenario the new physics scale characterized by the right-handed $W$ boson mass has been severely constrained by the LHC data~\cite{ATLAS:2023cjo}. This motivates us to consider other kinds of UV scenarios with chiral enhancement at low energies.

The UV completions of the dim-9 quark-lepton operators responsible for $\onbb$ decay have been systematically classified at tree level~\cite{Bonnet:2012kh} and one-loop level~\cite{Chen:2021rcv}. 
With the development of the EFTs, new ways of constructing the UV completions of the effective operators in the SMEFT have been proposed~\cite{Li:2022abx}, which have been used to obtain the complete UV resonances at tree level up to the dim-7 level~\cite{Li:2023cwy} and the dim-8 level~\cite{Li:2023pfw}. Following these, we find that there are certain tree-level UV completions that have not been discussed in Ref.~\cite{Bonnet:2012kh}.

In this work, rather than investigating all UV completions~\footnote{Although we use the terminologies `` UV completion'' and ``UV model'', the models we discuss might be embedded into more fundamental UV theories, see Ref.~\cite{Dvali:2023snt} for example for the related topic. In this sense, they are actually simplified models. } of the dim-9 SMEFT operators for $\onbb$ decay, we will focus on the UV resonances that give rise to the chiral enhancement while not severely constrained experimentally. Technically, we
investigate possible ``two-step'' UV relations of the dim-9 quark-lepton operators, namely from LEFT to the SMEFT and then to the UV models, and focus on the operators that lead to chiral enhancement at the hadronic scale.

We find that leptoquarks (LQs) are good candidates for the UV completions of the operators $O_4 \bar{e}_L e_L^c$ and $O_4 \bar{e}_R e_R^c$, which are weakly constrained compared to the right-handed $W$ boson~\cite{ATLAS:2020dsk} with the mass $m_{\rm LQ} \geq 1.8\tev$.
These UV resonances could be probed by both future tonne-scale $0\nu\beta\beta$-decay experiments indirectly and the current and future LHC experiments directly through searching for lepton number violation, which is important to 
 uncover the mechanisms for $\onbb$ decay. Previous studies of $\Delta L =2$ lepton number violation induced by LQs at the LHC can be found in Refs.~\cite{Hirsch:2013dla,Helo:2013ika,Gonzalez:2016ztm,Graesser:2022nkv}.

The remainder of the paper is organized as follows. In the next section, the effective operators for $\onbb$ decay in the SMEFT and LEFT with chiral enhancement are studied. In Sec.~\ref{sec:UV-completion}, possible UV completions of the SMEFT operators with LQs are obtained. In Sec.~\ref{sec:0vbb}, the half-life of $\onbb$ decay expressed in terms of the Wilson coefficients of the LEFT operators is given and the reach to the LNV scale is estimated. In Sec.~\ref{sec:lhc}, LHC searches for the UV resonances are investigated. In Sec.~\ref{sec:result}, the sensitivities to the UV model in the $\onbb$-decay and LHC searches are combined and discussed. We conclude in Sec.~\ref{sec:conclusion}.

\section{Effective operators for \tf{$\onbb$}{0vbb} decay}
\label{sec:operator}

In this work, we will study the effective operators that give rise to chirally enhanced contributions to $0\nu\beta\beta$ decay. In the LEFT below the EW scale, the $\Delta L =2$ quark-lepton interactions responsible for $\onbb$ decay are expressed as~\cite{Graesser:2016bpz,Cirigliano:2018yza}
\begin{equation}
	\label{eq:left}
	\begin{aligned}
		\mathcal{L}_{\rm LEFT}^ {(9)} \supset \frac{1}{v^5} \sum_{i=1}^{8} \left( C_{i L}^{(9)} \bar{e}_L e_L^c  +  C_{i R}^{(9)} \bar{e}_R e_R^c  \right) O_i  + \cdots \;,
	\end{aligned}
	\end{equation}
where the quark operators $O_i$ are explicitly given in Ref.~\cite{Cirigliano:2018yza}, $v=246\gev$, and the dots depict the terms with different lepton bilinears. 

We consider the scenario where only the dim-9 operator $O_{4X}^{(9)} \equiv O_4 \bar e_X e_X^c$ with $X=L$ or $R$ is generated directly after the electroweak symmetry breaking (EWSB). The quark operator is defined as~\cite{Graesser:2016bpz,Cirigliano:2018yza}
\begin{equation}
\begin{aligned}
\label{eq:O4-LEFT}
O_4 & =\bar{q}_L^\alpha \gamma_\mu \tau^{+} q_L^\alpha \bar{q}_R^\beta \gamma^\mu \tau^{+} q_R^\beta\;,  
\end{aligned}
\end{equation}
where $q_{L/R} = (u,d)_{L/R}^T$ are the left- and right-handed isospins, 
$\tau^+=\left(\tau_1 + i \tau_2\right) / 2$ with $\tau_i$ the Pauli matrices, and $\alpha, \beta$ are color indices. 
Hereafter, for a chiral fermion field $\psi$, we use the notations
$\psi^c \equiv  C \bar\psi^T$ and and $\bar\psi^c = \psi^T C$ with $C$ the charge-conjugation matrix.

The operator $O_{4X}^{(9)}$ would mix with $O_{5X}^{(9)} \equiv O_5 \bar e_X e_X^c$ due to the QCD renormalization group evolution (RGE), where the quark operator
\begin{align}
O_5 & =\bar{q}_L^\alpha \gamma_\mu \tau^{+} q_L^\beta \bar{q}_R^\beta \gamma^\mu \tau^{+} q_R^\alpha\;.
\end{align}
The corresponding RGE of the Wilson coefficients is~\cite{Buras:2000if,Buras:2001ra,Cirigliano:2018yza}
\begin{align}
\label{eq:matching2}
	\frac{d}{d\ln{\mu}}\begin{pmatrix}
		C_{4X}^{(9)}\\C_{5X}^{(9)}
	\end{pmatrix}=\frac{\alpha_s}{2\pi}\begin{pmatrix}
		   1&0\\
		-3&-8
	\end{pmatrix}\begin{pmatrix}
		C_{4X}^{(9)}\\C_{5X}^{(9)}
	\end{pmatrix}\;,
\end{align}
where $\alpha_s$ is the strong coupling. From the scale $ m_W = 80.4\gev$ to the hadronic scale $\Lambda_H = 2\gev$, we obtain
\begin{align}
\label{matching3}
    \begin{pmatrix}
         C_{4X}^{(9)}(\Lambda_H)\\C_{5X}^{(9)}(\Lambda_H)
    \end{pmatrix}=\begin{pmatrix}
        0.90 & 0\\
      0.45& 2.3
    \end{pmatrix}
    \begin{pmatrix}
         C_{4X}^{(9)}(m_W)\\C_{5X}^{(9)}(m_W)
    \end{pmatrix}\;.
\end{align}

Both the quark-lepton operators $O_{4X}^{(9)}$ and $O_{5X}^{(9)}$ can be mapped to the $\Delta L =2$ hadron-leptpn operators at the hadronic scale with chiral enhancement in chiral effective field theory, see Refs.~\cite{Prezeau:2003xn,Graesser:2016bpz,Cirigliano:2018yza} for details.
From Eq.~\eqref{matching3}, we can see that a non-zero $C_{5X}^{(9)}$ is induced at the hadronic scale although it is vanishing at the scale $\mu = m_W$. The half-life of $\onbb$ decay including $C_{4X}^{(9)}$ and $C_{5X}^{(9)}$ will be given in Sec.~\ref{sec:0vbb}.

Above the EW scale, 
effective operators in the SMEFT written as 
\begin{align}
\mathcal{L}_{{\rm SMEFT}}^{(d)} =  \sum_{i} \dfrac{\mathcal{C}_i^{(d)}}{\Lambda^{d-4}} \mathcal{O}_i^{(d)}\;,
\end{align}
where $\mathcal{C}_i^{(d)}$ are the Wilson coefficients, and $\Lambda$ is the UV scale.
In order to generate $\Delta L =2$ lepton number violation, the mass dimension $d$ should be an odd number. For $d=5$, the Weinberg operator~\cite{Weinberg:1979sa} arises, which is irrelevant to $O_{4X}^{(9)}$. For $d=7,9$, the complete bases of operators
have been obtained in Refs.~\cite{Lehman:2014jma,Liao:2016hru,Liao:2020jmn,Li:2020xlh}. 

At the dim-7 level~\cite{Lehman:2014jma, Liao:2016hru}, there is only one SMEFT operator that is related to  $O_{4X}^{(9)}$~\cite{Scholer:2023bnn}, 
\begin{align}
\label{eq:smeft-dim7}
\mathcal O_{\bar d u LLD}^{(7)} &=\epsilon^{i j}\left(\bar{d}_R \gamma^\mu u_R\right)\left(\bar{L}_i^c i D_\mu L_j\right)\;,
\end{align}
where $L=\left(\nu_e, e\right)_L^T$ is the SM lepton doublet with the flavor index omitted,  and $D_\mu$ is the covariant derivative,  $\epsilon \equiv i\tau_2 $ is the antisymmetric tensor with $\tau_2$ the second Pauli matrix, and $D_\mu L_j \equiv \left(D_\mu L\right)_j$.

At the dim-9 level, there are four relevant SMEFT operators~\cite{Scholer:2023bnn}, which are expressed as~\cite{Li:2020xlh}
\begin{align}
\label{eq:smeft-dim9}
\mathcal O_1^{(9)} &=\epsilon^{i j}\left(\bar{d}_{R} \gamma^\mu e_R\right)\left(\bar{u}^c_R e_R\right)H_j D_\mu H_i\;,\nonumber\\
\mathcal O_2^{(9)} &=\epsilon ^{ik} \left(\bar{d}_{R} L_{j}\right) \left(\bar{L}_{i}^c \gamma ^{\mu } u_{R}\right) H^{\dagger}{}^{j} D_\mu H_k\;,\nonumber\\
\mathcal O_3^{(9)} &=\epsilon ^{ij} \left(\bar{d}_{R} \gamma^{\mu } u_{R}\right) \left(\bar{L}_i^c D_\mu L_j\right) H_{k} H^{\dagger}{}^{k}\;,\nonumber\\
\mathcal{O}_{4}^{(9)}&=\epsilon^{ik}(\bar{u}_R^\alpha Q^{\beta}_j )(\bar{L}^jd_R^{\alpha})(\bar{L}_iQ_{k}^{\beta c})\;.
\end{align}
Here,  $D_\mu H_i \equiv \left(D_\mu H\right)_i$ and $H^{\dagger}{}^{j} \equiv  (H^{\dagger})^j$ are used for brevity, and $Q$ is the left-handed quark doublet with $\alpha$, $\beta$ the color indices.

Notice that the quark and lepton fields are contracted with each other in the operators $\mathcal{O}_1^{(9)}$,   $\mathcal{O}_2^{(9)}$ and $\mathcal{O}_4^{(9)}$.
In order to match with $O_{4X}^{(9)} $, we use the Fierz relations derived in Ref.~\cite{Liao:2016hru}, 
and express them as
\begin{align}
\mathcal{O}_1^{(9)}&=-\dfrac{1}{2} \epsilon^{i j}(\bar{d}_R\gamma^{\mu}u_R)(\bar{e}_R^c e_R) H_j D_\mu H_i \;,\\
\mathcal O_2^{(9)} 
&=-\epsilon ^{ik} \left(\bar{d}_{R} L_{i}\right) \left(\bar{L}_{j}^c \gamma ^{\mu } u_{R}\right) H^{\dagger}{}^{j} D_\mu H_{k}\nonumber\\
&\quad  -\epsilon ^{ik} \left(\bar{L}_i^c L_{j}\right) \left(\bar{d}_R \gamma ^{\mu } u_{R}\right) H^{\dagger}{}^{j} D_\mu H_{k}\;,\label{eq:Firez-O2}\\
 \label{eq:Firez-O4}
\mathcal O_4^{(9)} 
 &=-\frac{1}{2}
 \epsilon^{ik}(\bar{u}_R^\alpha \gamma^{\mu}d_R^{\alpha} )(\bar{L}^j \gamma_{\mu} Q^{\beta }_j)(\bar{L}_iQ_{k}^{\beta c})\nonumber\\
&=\frac{1}{2}
\epsilon^{ik}(\bar{u}_R^\alpha \gamma^{\mu}d_R^{\alpha} )(\bar{L}_i \gamma_{\mu} Q^{\beta}_j)(\bar{L}^jQ_{k}^{\beta c})\nonumber\\
&\quad  +\frac{1}{2}
\epsilon^{ik}(\bar{u}_R^\alpha \gamma^{\mu}d_R^{\alpha} )(\bar{Q}_{k}^{\beta } \gamma_{\mu} Q^{\beta}_j)(\bar{L}_i L^{jc} )\;.
\end{align}

The isospin indices $i=j$ are further required in Eqs.~\eqref{eq:Firez-O2}~\eqref{eq:Firez-O4},
so that $\mathcal{O}_2^{(9)}$ and $\mathcal{O}_4^{(9)}$ can be converted into the operators with separate quark and lepton bilinears. As a result, we can obtain that the Wilson coefficients of the SMEFT operators in Eqs.~\eqref{eq:smeft-dim7}~\eqref{eq:smeft-dim9} remain the same above the EW scale since the vector quark current does not evolve in QCD~\cite{Cirigliano:2018yza}. 

After the EWSB, the Wilson coefficients of the SMEFT operators are matched to $C_{4X}^{(9)}$. The matching conditions
at the scale $\mu=m_W $ are~\cite{Scholer:2023bnn}
 \begin{align}
 \label{eq:matching}
{C_{4R}^{(9)}}(m_W)&=\frac{i}{2} V_{u d} \frac{v^5}{\Lambda^5}{\mathcal C}_1^{(9)*}\;,\\
{C_{4L}^{(9)}}(m_W)&=  \frac{i}{2}V_{u d}\frac{v^5}{\Lambda^5} \left( \mathcal{C}_2^{(9)*} + 2\mathcal{C}_3^{(9)*} \right)\nonumber\\
&\quad + \frac{1}{4}\frac{v^5}{\Lambda^5}\mathcal{C}_4^{(9)}\label{leftmatching} -2  V_{u d}\frac{v^3}{\Lambda^3} \mathcal{C}_{\bar{d}uLLD}^{(7)*}\;,
\end{align}
where $V_{ud} = 0.974$ denotes the quark mixing in the SM.  We can see that the contributions of the dim-9 SMEFT operators in Eq.~\eqref{eq:smeft-dim9} to the Wilson coefficients of the LEFT operators $\mathcal{O}_{4X}^{(9)}$ $X=L,R$ are proportional to $v^5/\Lambda^5$~\footnote{The contribution of $\mathcal{O}_4^{(9)}$ is comparable to the other three dim-9 SMEFT operators, since for the latter both Higgs fields develop the vevs that cancel $m_W^2$ in the propagator in the matching. 
}, while that of the dim-7 operator in Eq.~\eqref{eq:smeft-dim7} is proportional to $v^3/\Lambda^3$.

\section{UV completion}
\label{sec:UV-completion}

In this section, we will investigate possible UV completions of the SMEFT operators discussed in Sec.~\ref{sec:operator}. It is noted that there is no tree-level UV completion of the dim-7 SMEFT operator $\mathcal O_{\bar d u LLD}^{(7)}$~\cite{Li:2022abx}. The Wilson coefficient $\mathcal C_{\bar d u LLD}^{(7)}$ obtained by integrating out heavy fields at one-loop level is suppressed by the loop factor $ 1/(16\pi^2)$, thus it is expected to be comparable to the Wilson coefficients of the dim-9 SMEFT operators $\mathcal{O}_{i}^{(9)}$ $\left(i=1,\cdots,4\right)$ if they are obtained at tree-level in UV models with the LNV scale $\Lambda \sim 4\pi v$.

Moreover, the dim-9 SMEFT operator $\mathcal{O}_1^{(9)}$ can be realized in the LRSM from the exchange of both left- and right-handed $W$ bosons~\cite{Prezeau:2003xn,Cirigliano:2018yza,Li:2020flq}. As mentioned in Sec.~\ref{sec:intro}, such contribution is suppressed by the right-handed $W$ boson mass. Besides, $\mathcal{O}_4^{(9)}$ could originate from the exchange of charged scalars, which was considered in the LRSM~\cite{Mohapatra:1981pm}. This is easily verified if one converts $\mathcal O_4^{(9)}$ into the following SMEFT operator in the basis of Ref.~\cite{Liao:2020jmn} 
\begin{align}
\mathcal{O}^{(9)}_{dQQuLL2} =  \epsilon^{ij} \left( \bar{d}^\alpha Q_{i}^\beta \right) \left( \bar{Q}^\beta u^\alpha \right) \left( \bar{L}^c L_j \right)
\end{align}
using the Fierz identities. 
However, such a physical scenario is severely constrained by the charged lepton flavor violating searches~\cite{Tello:2010am,Li:2020flq}.

In the following, we will study possible UV completions with the LQs that are weakly constrained. In Tab.~\ref{tab:fields}, the new fields we introduce are listed, and their interactions with the SM fields are given for each case.

\begin{table*}[ht]
\renewcommand\arraystretch{1.7}
\caption{The new fields with the corresponding quantum numbers $(X, Y,Z)$  under the $SU(3)_C$, $SU(2)_L$ and $U(1)_Y$ gauge groups are shown. We label the masses of scalar LQs: $\tilde{R}_2$, $\bar{S}_1$, ${S}_1$, vector LQs: $U_1$, $\tilde{V}_2$, vector-like fermions: $\Psi = (\Psi_L,\Psi_R)$, $E^\prime = (E_L^\prime, E_R^\prime)$, $d^\prime = (d^\prime_L,d^\prime_R)$, and singlet scalar: $S$ as $m_R$, $m_{\bar{S}_1}$,$m_{{S}_1} $, $m_U$,  $m_V$, $m_\Psi$, $m_{E^\prime}$, $m_{d^\prime}$ and $m_S$, respectively. The vector-like fermions are introduced to make the models anomaly-free. We follow the notations of Ref.~\cite{Buchmuller:1986zs}  for LQs, see also Ref.~\cite{Dorsner:2016wpm}.  }
\begin{center}
\begin{tabular}{c|c|c|c|c}
\hline
\hline
operator    & \multicolumn{2}{c|}{leptoquark(s)}  & vector-like fermions &  singlet scalar \\ \hline
$\mathcal{O}_1^{(9)}$ &  $\tilde{R}_2\in(3,2,1/6)$ & $U_1\in(3,1,2/3)$ & $\Psi_{L,R} \in(1,2,-1/2)$& / \\ \hline
$\mathcal{O}_2^{(9)}$ &  $\bar{S}_1\in(\bar{3},1,-2/3)$ & $\tilde{V}_2\in(\bar{3},2-1/6)$ & $E_{L,R}^\prime\in(1,1,-1)$ & / \\ \hline
$\mathcal{O}_3^{(9)}$ &  $\tilde{R}_2\in(3,2,1/6)$ & / & $\Psi_{L,R} \in(1,2-1/2)$ & $S\in(1,1,0)$ \\ \hline
$\mathcal{O}_4^{(9)}$ & $\tilde{R}_2\in(3,2,1/6)$ & $S_1\in(\bar{3},1,1/3)$ & $\Psi_{L,R} \in(1,2-1/2)$ & /  \\ \hline
$O_{\bar d u LLD}^{(7)}$ & $\tilde{V}_2\in(\bar{3},2-1/6)$ & / & $\Psi_{L,R} \in(1,2,-1/2)$, $d^\prime_{L,R} \in(3,1,-1/3)$ & $S\in(1,1,0)$ \\ \hline
\hline
\end{tabular}
\end{center}
\label{tab:fields}
\end{table*}

\subsection{Dim-9 SMEFT operator: \tf{$\mathcal O_1^{(9)}$}{O3}}

Firstly, we consider a UV model with the following interactions as the explicit completion of $\mathcal O_1^{(9)}$:  
\begin{align}
\label{eq:lag1}
\mathcal{L}&\supset \lambda_{ed} \left(\bar{d}_R \gamma_\mu e_R\right) U_1^\mu+ \lambda_{u\Psi} \tilde R_2^* \bar{u}^c_R \Psi_R\nn\\
		&\quad +\lambda_{DH} U_1^{\mu \dagger} \tilde R_2 \epsilon\left(iD_\mu H\right)+f_{\Psi e}\bar{\Psi}_L H e_R+ \text{h.c.}\;,
\end{align}
where ``$\text{h.c.}$'' represents the Hermitian conjugate terms. A scalar LQ field $\tilde R_2$ , a vector LQ field $U_1$ and vector-like fermions
$\Psi_L, \Psi_R$ with opposite chirality are introduced with the quantum number being specified in Tab.~\ref{tab:fields}. The vector-like fermion doublets are expressed as
\begin{equation}
\label{Dirac fermion}
	\Psi_L=\left(\begin{array}{c}
		N_L \\
		E_L
	\end{array}\right)\;, \quad \Psi_R=\left(\begin{array}{c}
		N_R \\
		E_R
	\end{array}\right)\;.
\end{equation}
The kinetic and mass terms are omitted here, and discussion on the mixing of lepton fields is given in Appendix~\ref{app:mixing}.

\begin{figure}[h]
	\centering
 	\includegraphics[width=0.7\linewidth]{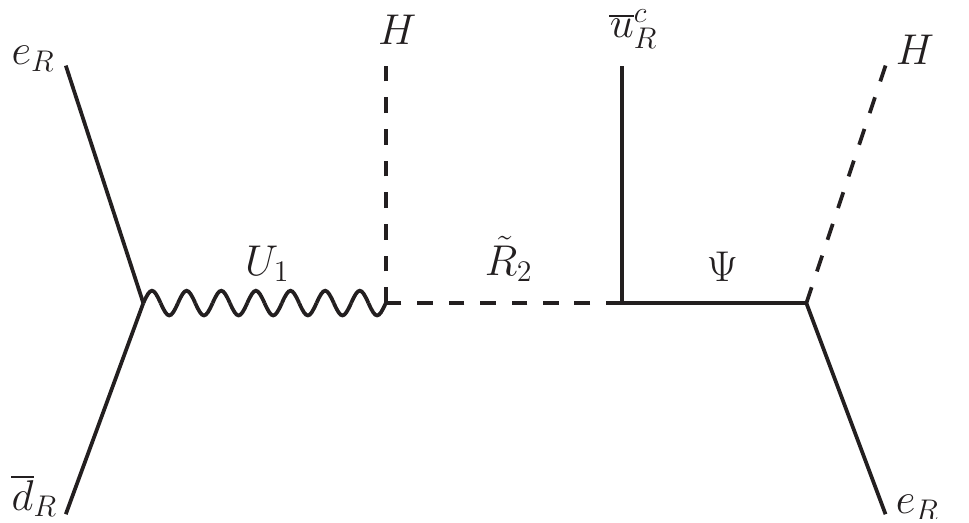}
	\caption{Feynman diagram for the UV completion of $\mathcal{O}_1^{(9)}$.
 }
	\label{fig:model1}
\end{figure}

After integrating out the heavy fields $U_1$, $\tilde{R}_2$ and $\Psi=(\Psi_L,\Psi_R)$ at tree level, we obtain the dim-9 SMEFT operator $\mathcal{O}_1^{(9)}$ (cf. Fig.~\ref{fig:model1})
with the Wilson coefficient
\begin{equation}
\label{eq:matching1}
	\frac{{\mathcal C}_{1}^{(9)}}{\Lambda^5}= i \frac{\lambda_{ed} \lambda_{u\Psi} \lambda_{DH} f_{\Psi e}}{m_U^2 m_R^2 m_{\Psi}}\;,
\end{equation}
where $m_U$, $m_R$ and $m_\Psi$ denote the masses of $U_1$, $\tilde{R}_2$ and $\Psi$, respectively. 

If all of the couplings in Eq.~\eqref{eq:lag1} are non-vanishing,  $\lambda_{ed}\lambda_{u\Psi}\lambda_{DH}f_{\Psi e}\neq 0$, there is $\Delta L =2$ lepton number violation, and {\it vice versa}.
To see it, following Ref.~\cite{Graesser:2022nkv} we consider a fictitious lepton number $U(1)_L$ under which the fields are charged and the couplings are treated as spurions to make the Lagrangian in Eq.~\eqref{eq:lag1} invariant. Let $q(U_1) = r$, $q(\tilde{R}_2) = s$, $q(\Psi) = t$, we have $q(\lambda_{ed}) = -(r+1)$, $q(\lambda_{u\Psi}) = s-t$, $q(\lambda_{DH}) = r-s$, $q(f_{\Psi e}) = t-1$, which implies that $q(\lambda_{ed} \lambda_{u\Psi} \lambda_{DH} f_{\Psi e}) = -2$.

\subsection{Dim-9 SMEFT operator: \tf{$\mathcal O_2^{(9)}$}{O3}}

Then we introduce a UV model as an explicit completion of $\mathcal{O}_2^{(9)}$ with the interactions as follows:
\begin{equation}
	\begin{aligned}
		\mathcal{L} & \supset f_{LE}(\bar{L}E_R^\prime)H +\lambda_{DH}(iD_{\mu}H)^{\dagger} \tilde V_2^{\mu}\bar{S}_1^{*}\\ 
  &+\lambda_{Ed}(\bar{E}_L^\prime d_R)\bar{S}_1+\lambda_{Lu}(\bar{L}\gamma_\mu u^c_R)\epsilon \tilde V_2^{\mu\dagger} + {\rm h.c.}\;,
	\end{aligned}
\end{equation}
where the new fields we introduce are vector-like fermions $E_L^\prime, E_R^\prime $, a scalar LQ $\bar{S}_1 $, a vector LQ $\tilde V_2 $, which are specified in Tab.~\ref{tab:fields}.

\begin{figure}[h]
	\centering
	\includegraphics[width=0.7\linewidth]{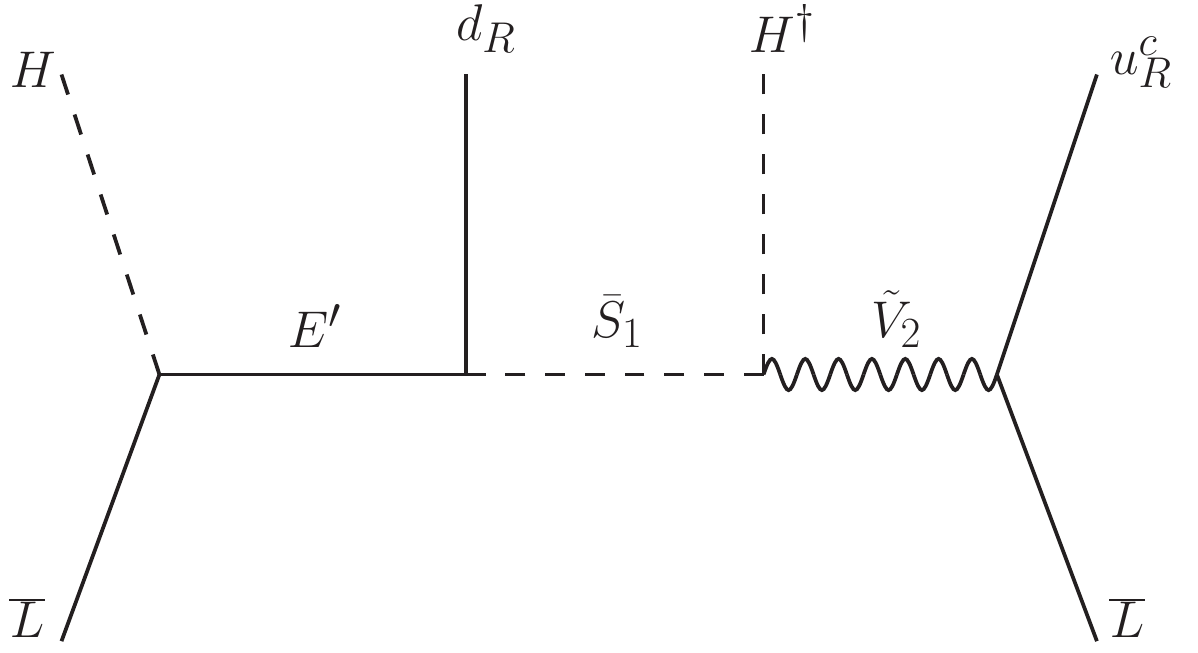}
	\caption{Feynman diagram for the UV completion of $\mathcal{O}_2^{(9)}$.  }
	\label{fig:model2}
\end{figure}

After integrating out the heavy fields at tree level (cf. Fig.~\ref{fig:model2}.), we can obtain the operator $\mathcal{O}_2^{(9)}$ with the Wilson coefficient
\begin{equation}
\label{eq:matching3}
\frac{{\mathcal C}_{2}^{(9)*}}{\Lambda^5}=-i  \frac{f_{LE}\lambda_{Ed} \lambda_{DH}\lambda_{Lu}}{m_{E^{\prime}}{m_{\bar{S}_1}^{2}}{m_{V}^2}}\;,
\end{equation}
The condition $f_{LE}\lambda_{Ed} \lambda_{DH}\lambda_{Lu}\neq 0$ implies lepton number violation as the case of  $\mathcal{O}_1^{(9)}$.

\subsection{Dim-9 SMEFT operator: \tf{$\mathcal O_3^{(9)}$}{O3}}

The operators $\mathcal{O}_1^{(9)}$ and $\mathcal{O}_2^{(9)}$ involve the covariant derivative on the Higgs doublet $D_\mu H$, which can be generated directly after integrating out the heavy fields. The situation of the operator $\mathcal{O}_3^{(9)}$ is different, as the mass dimension of $D_\mu L$ is $5/2$, so that it can only emerge from the kinematic terms of fermions. 

An example UV completion of $\mathcal{O}_3^{(9)}$ is given as follows
\begin{align}
\label{eq:lag3}
	\mathcal{L}&\supset \lambda_{Ld}(\bar{L}d_R)\epsilon\tilde{R}_2^*+\lambda_{u\Psi}(\bar{\Psi}_Ru_R^c) \tilde{R}_2\nonumber\\
	&+f_{L\Psi}(\bar{L}\Psi_R)S+\mu(H^{\dagger}H)S^*\;,
\end{align}
where a scalar LQ $\tilde{R}_2$, vector-like fermions $\Psi$ and a real singlet scalar $S$ are introduced, which are listed in Tab.~\ref{tab:fields}.
\begin{figure}[h]
	\centering
	\includegraphics[width=0.7\linewidth]{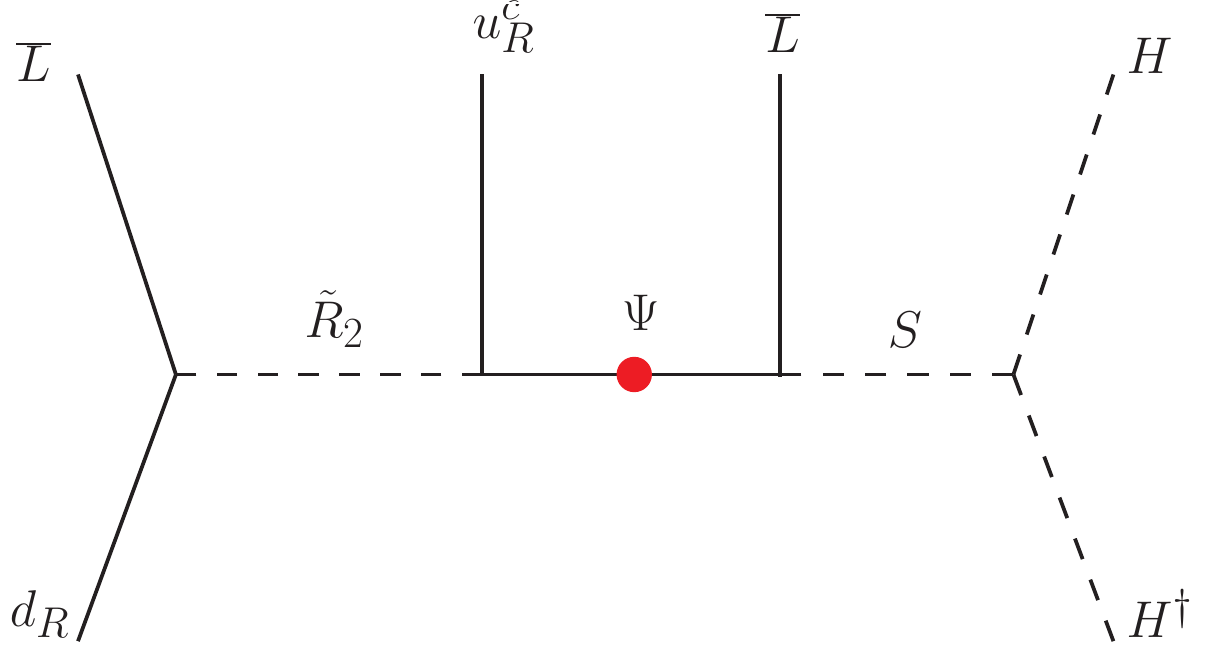}
	\caption{Feynman diagram for the UV completion of $\mathcal{O}_3^{(9)}$. The red dot labels the insertion of the covariant derivative.}
	\label{fig:model3}
\end{figure}

After integrating out the heavy fields at tree level (cf. Fig.~\ref{fig:model3}), we can obtain the following effective interactions
\begin{align}
\label{eq:CD3}
\mathcal{L}_{\text{eff}}\supset  & -\frac{\lambda_{Ld}\lambda_{u\Psi}f_{L\Psi}\mu}{m_{{R_2}}^2m_{\Psi}^2m_S^2} \epsilon_{ij}	\left[(\bar{L}^i d_R)(\bar{L}^j\gamma_{\mu}iD^{\mu}u^c_R)(H^{\dagger}H)\right.\nonumber\\
&\qquad\qquad +  (\bar{L}^i iD^{\mu} d_R)(\bar{L}^j\gamma_{\mu} u^c_R)(H^{\dagger}H)\nonumber\\
& \qquad\qquad + \left.(iD^{\mu}\bar{L}^i d_R)(\bar{L}^j\gamma_{\mu} u^c_R)(H^{\dagger}H)
\right]\;.
\end{align}

The covariant derivatives of the SM fields in Eq.~\eqref{eq:CD3} is tracked from the kinematic terms of the vector-like fermions  $\Psi$.
The detail is given in Appendix~\ref{app:CDE}. 
Since the covariant derivatives $D^{\mu}u^c_R$ and $D^{\mu}d_R$ do not involve the $W$ boson and would not generate $0\nu\beta\beta$ decay at lower energies, only the third term 
is relevant. We label it as $\mathcal{O}_3^{(9)\prime}$ with the coefficient 
 \begin{align}
 \label{match3}
\frac{{\mathcal {C}}_3^{(9)\prime}}{\Lambda^5}
 =-\frac{\lambda_{Ld}\lambda_{u\Psi}f_{L\Psi}\mu}{m_{R}^2m_{\Psi}^2m_S^2}\;.
 \end{align}

Using the Fierz relations derived in Ref.~\cite{Liao:2016hru}, we can convert $\mathcal{O}_3^{(9)\prime}$ into $\mathcal{O}_3^{(9)}$:
 
\begin{align}
\label{03}
\mathcal{O}_3^{(9)\prime} = -i\mathcal{O}_3^{(9)\dagger} + \epsilon_{ij}(\bar{u}_R \gamma_{\mu}iD^{\mu}L^i{}^c)(\bar{L}^jd_R)(H^{\dagger}H)\;,
\end{align}
Thus the Wilson coefficient~\footnote{It is noted that in Eq.~\eqref{03}, the second term, which is redundant by using the equation of motion, also contributes to the Wilson coefficient of the LEFT operator $O_{4L}^{(9)}$. Thus the matching condition at the scale $\mu = m_W$ is 
 \begin{align}
 \label{o3prime}
 {C_{4L}^{(9)}}(m_W)
 =\frac{1}{2}V_{u d} \frac{v^5}{\Lambda^5}{\mathcal C}_3^{(9)\prime}(m_W)\;.
 \end{align}
 
}
\begin{align}
\frac{{\mathcal {C}}_3^{(9)\prime}}{\Lambda^5} = i \frac{{\mathcal {C}}_3^{(9)*}}{\Lambda^5}\;.
\end{align}
Similarly, the condition of lepton number violation is $\lambda_{Ld}\lambda_{u\Psi}f_{L\Psi}\mu\neq 0$.

\subsection{Dim-9 SMEFT operator: \tf{$\mathcal O_4^{(9)}$}{O4}}

Besides the LRSM mentioned before, the six-fermion operator $\mathcal O_4^{(9)}$ can also be realized in the following UV model:
\begin{align}
    \mathcal{L}\supset &\lambda_{Ld}(\overline{L}d_R)\epsilon\tilde{R}_2^*+\lambda_{u\Psi}(\bar{\Psi}_R u_R^c)\tilde{R}_2\nonumber\\
    &+f_{\Psi S}S_1(\overline{Q}^c\epsilon \Psi_L)+f_{LQ}(\overline{L}\epsilon Q^c)S_1^* + {\rm h.c.}\;,
\end{align}
where two scalar leptoquarks $\tilde{R}_2$, $S_1$, and vector-like fermions
$\Psi_L, \Psi_R$ with opposite chirality are introduced with the quantum number being specified in Tab.~\ref{tab:fields}. The similar model was introduced in Table 8 of Ref.~\cite{Bonnet:2012kh} as the UV completion of the LEFT operator $O_{4L}^{(9)}$.

\begin{figure}[h]
 	\centering
 	\includegraphics[width=0.7\linewidth]{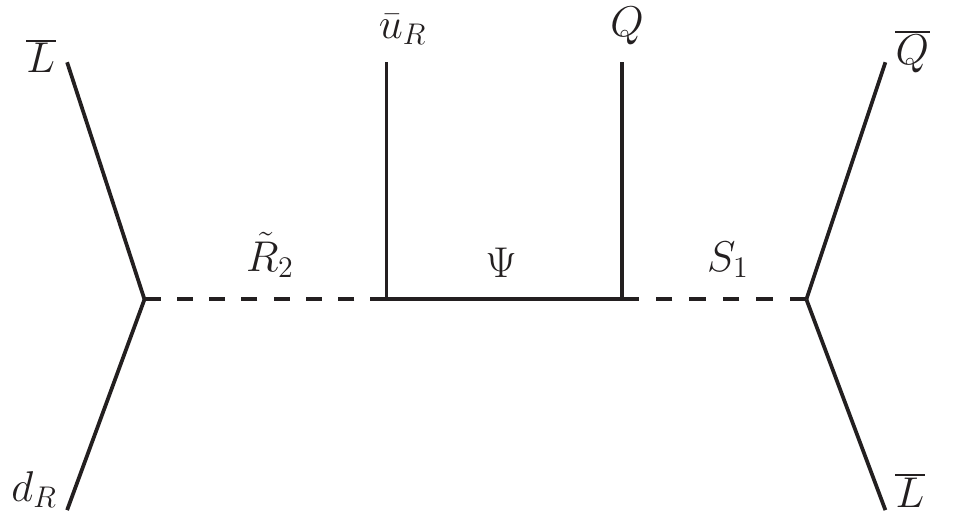}
 	\caption{Feynman diagram for UV completion of $\mathcal{O}_4^{(9)}$.  }
	\label{fig:model4}
 \end{figure}

 After integrating out the heavy fields at the tree level (cf. Fig.~\ref{fig:model4}), we have
 \begin{align}
 \mathcal{C}_4^{(9)}=-\frac{\lambda_{Ld}\lambda_{u\Psi} f_{\Psi S}f_{LQ}}{m_{R}^2m_{\Psi}m_{S_1}^2} \;.
 \end{align}
and the condition for lepton number is $\lambda_{Ld}\lambda_{u\Psi} f_{\Psi S}f_{LQ} \neq 0$. 

\subsection{Dim-7 SMEFT operator: \tf{$\mathcal O_{\bar d u LLD}^{(7)}$}{OduLLD}}

\begin{figure}[h]
	\centering
 	\includegraphics[width=0.8\linewidth]{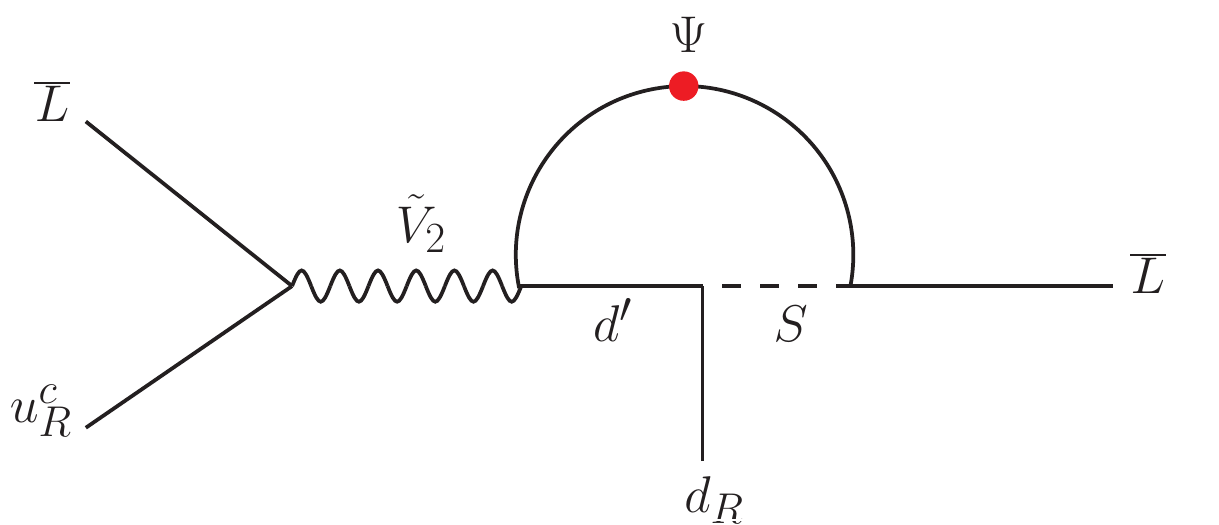}
	\caption{Feynman diagram for the UV completion of $\mathcal O_{\bar d u LLD}^{(7)}$. The red dot labels the insertion of the covariant derivative.
 }
 \label{fig:model5}
\end{figure}

As obtained in Ref.~\cite{Li:2022abx}, there is no tree-level UV completion of the dim-7 SMEFT operator $\mathcal O_{\bar d u LLD}^{(7)}$. A possible one-loop realization 
is described by the following Lagrangian:
\begin{align}
    \mathcal{L}& \supset \lambda_{Lu}(\bar{L}\gamma^{\mu}u_R^c)\epsilon \tilde{V}_{2\mu}^{\dagger}+\lambda_{\Psi d^\prime}(\bar{\Psi}_R\gamma^{\mu}d^\prime_R)\tilde{V}_{2\mu}\nonumber\\
    &+f_{d^\prime d}(\bar{d}^\prime_L d_R)S+f_{L\Psi}(\bar{L}\Psi_R)S + {\rm h.c.}\;,
\end{align}
where we have introduced a vector LQ $\tilde{V}_2\in(\bar{3},2,-1/6)$, vector-like fermions $\Psi_{L,R}\in(1,2,-1/2)$ and $d_{L,R}^\prime \in(3,1,-1/3)$, and a real singlet scalar $S\in(1,1,0)$. 
 The UV completion of the dim-7 SMEFT operator $\mathcal O_{\bar d u LLD}^{(7)}$ is shown in Fig.~\ref{fig:model5}. The covariant derivative is also tracked from the kinetic term of $\Psi$ similar to the dim-9 SMEFT operator $\mathcal{O}_3^{(9)}$, see more details in Appendix~\ref{app:CDE}.

Using the naive dimensional analysis~\cite{Weinberg:1978kz,Manohar:1983md,Gavela:2016bzc}, we can estimate the Wilson coefficient as
\begin{align}
\mathcal{C}_{\bar d u LLD}^{(7)} \sim \dfrac{1}{(4\pi)^2} \lambda_{Lu} \lambda_{\Psi d^\prime} f_{d^\prime d} f_{L\Psi}\;.
\end{align}
Similarly, $\lambda_{Lu} \lambda_{\Psi d^\prime} f_{d^\prime d} f_{L\Psi}\neq 0$ implies lepton number violation.

We have some general discussions on the UV models introduced. 
Different from the previous studies, which either match the known UV theories (such as the LRSM) with the effective operators or construct the UV models based on the topologies of the LEFT operators, our approach in two steps enables new UV completions, which have not been investigated before. Besides, in all of these UV models, the neutrino masses firstly arise at $n$-loop level with $n=2,3,4$, which are negligible as we show in Appendix~\ref{app:neutrino-mass}. As we have discussed,  the contributions of dim-7 and dim-9 SMEFT operators to $\onbb$ decay are comparable, thus in order to diagnose the mechanism of $\onbb$ decay we need other complementary probes, and to search for the UV resonances in the UV models. 

Thus in the following sections, we will investigate the sensitivities to the mass and couplings of the UV resonances in $\onbb$ decay and at the LHC. For simplicity, we will assume that all couplings are real and positive.

\section{\tf{$0\nu\beta\beta$}{0vbb} decay }
\label{sec:0vbb}

In the EFT framework, the inverse half-life of $\onbb$ decay can be expressed as~\cite{Cirigliano:2018yza}
\begin{align}
\label{29}
\left(T_{1/2}^{0\nu}\right)^{-1} = g_A^4 &\left[G_{01}\left(\left|\mathcal{A}_L\right|^2+\left|\mathcal{A}_R\right|^2\right)\right.\nn\\
&\left.-2\left(G_{01}-G_{04}\right) \operatorname{Re} \mathcal{A}_L^* \mathcal{A}_R\right]\;,
\end{align}
and the amplitude is
\begin{align}
\label{eq:amplitude}
\mathcal A_X &= \dfrac{1}{2 m_e v} C_{\pi\pi X}^{(9)} \sum_{i=GT,T}\left( \dfrac{1}{2} M_{i,sd}^{AP} +M_{i,sd}^{PP}  \right)\;,\\
C_{\pi\pi X}^{(9)} &= - g_4^{\pi\pi} C_{4X}^{(9)} - g_5^{\pi\pi} C_{5X}^{(9)}\;,
\end{align}
where $X=L,R$,  $g_A=1.27$, and the Wilson coefficients  $C_{4X}^{(9)}$ and $C_{5X}^{(9)}$ are evaluated at $\mu=2\gev$. The low energy constants $g_4^{\pi\pi} = -1.9\gev^2$ and $g_5^{\pi\pi} = -8.0\gev^2$~\cite{Nicholson:2018mwc}.
For $^{136}$Xe, the phase-space factor $G_{01} = 1.5\times 10^{-15}$~year$^{-1}$, and the nuclear matrix elements using the quasi-particle random-phase approximation~\cite{Hyvarinen:2015bda,Cirigliano:2018yza}:
\begin{align}
M_{GT,sd}^{AP} &= -2.8\;,&  M_{GT,sd}^{PP} = 1.06\;,\nn\\
M_{T,sd}^{AP} &= -0.92\;,&  M_{T,sd}^{PP} = 0.36\;,
\end{align}

It should be noted that the pion-exchange neutrino potential scales as $1/{\bf q}^2$ at large $|{\bf q}|$ with ${\bf q}$ being the momentum transfer~\cite{Cirigliano:2018yza}. As a result, the amplitude of $\onbb$ decay is UV divergent~\cite{Cirigliano:2018hja}.  To absorb this divergence, the short-range $N NNN e e$ contact interaction needs to be promoted to leading order~\cite{Cirigliano:2018hja,Cirigliano:2019vdj}, which is of the same order as non-derivative $\pi\pi ee$ interaction~\cite{Cirigliano:2018yza}. 
 In this work, we use the amplitude $\mathcal{A}_X$ in Eq.~\eqref{eq:amplitude} to provide an estimate of the $\onbb$-decay rate.

 \begin{figure}[!htb]
\centering
\includegraphics[width=0.28\textwidth]{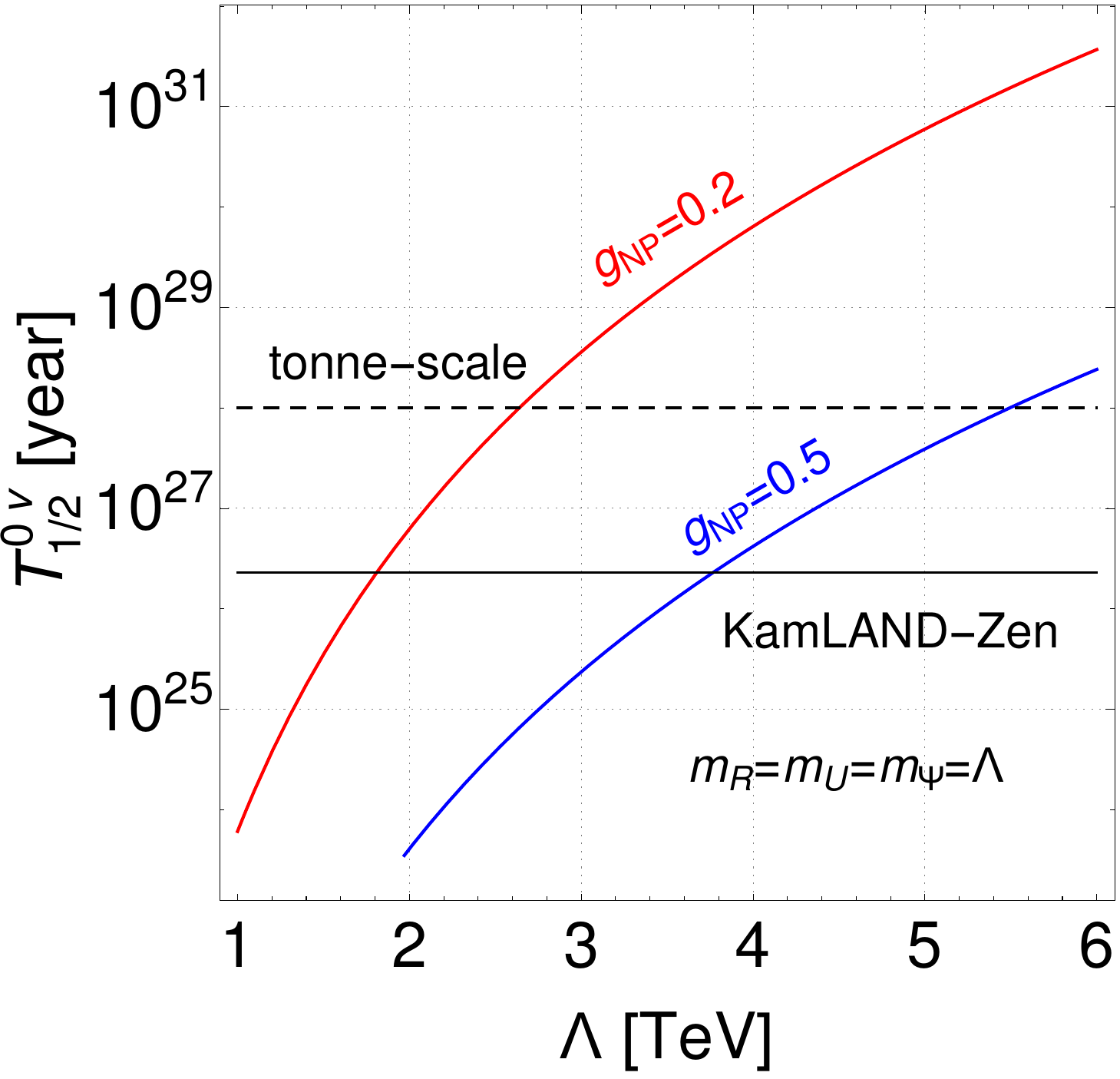}
\caption{Sensitivities to the LNV scale $\Lambda $ in the $\onbb$-decay experiments for the new physics couplings $g_{\rm NP} = 0.2$ (red) or 0.5 (blue). The solid and dashed black lines correspond to limits given by the KamLAND-Zen and future tonne-scale experiments, respectively. 
}
\label{fig:DBD_estimate}
\end{figure}
 
Currently, the most stringent constraint on the $\onbb$ decay is given by the KamLAND-Zen experiment, $T_{1/2}^{0\nu} > 2.3\times 10^{26}$~year at 90\% confidence level (C.L.)~\cite{KamLAND-Zen:2022tow}. The future tonne-scale experiments are expected to improve the sensitivity to the half-life by about two orders of magnitude $\sim 10^{28}$~year, see Ref.~\cite{Adams:2022jwx} and references therein. 

As an estimate, we consider the sensitivities to the LNV scale $\Lambda \equiv m_R = m_U = m_\Psi$ assuming two benchmark values of the new physics couplings $g_{\rm NP} \equiv \lambda_{ed} = \lambda_{u\Psi} = \lambda_{DH} = f_{\Psi}$ in the UV model for $\mathcal{O}_1^{(9)}$ in KamLAND-Zen and future tonne-scale $\onbb$-decay experiments. From Fig.~\ref{fig:DBD_estimate}, we find that the ongoing $\onbb$-decay experiments are able to reach the LNV scale $\Lambda \sim 2-3\tev$ or $4-5\tev$ for $g_{\rm NP}=0.2$ or 0.5, respectively.  
Given the existing constraints on the masses and couplings~\cite{Schmaltz:2018nls,Crivellin:2021egp,ATLAS:2020dsk}, we obtain that the sensitivities to the masses of UV resonances could be probed directly in the LHC searches.

\section{LHC searches}
\label{sec:lhc}

In this section, we will study the LHC searches for the UV resonances in the models discussed in Sec.~\ref{sec:UV-completion}.
The LQs are mainly produced in pairs, which subsequently decay into the SM quarks, leptons, or other new particles. Due to the lepton number violation, we can achieve the same-sign dilepton (SSDL) signature with a pair of same-sign electrons and at least two jets in the final state. For previous studies of lepton number violation in the final state of muon(s) in other contexts, see Refs.~\cite{Fuks:2020att,Fuks:2020zbm,ATLAS:2023tkz,CMS:2022hvh,Babu:2022ycv}.

\begin{figure}[!htb] 
\centering
\includegraphics[width=0.28\textwidth]{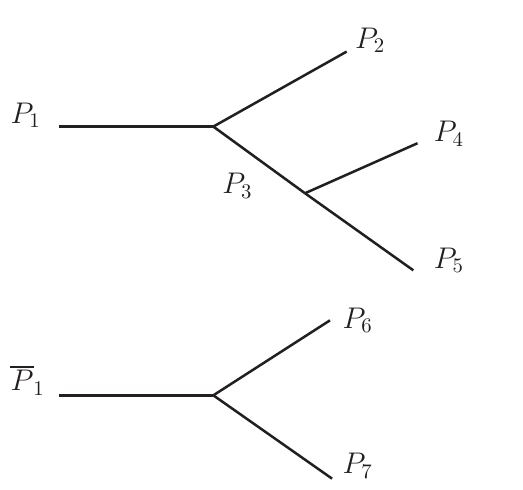}	
\caption{Diagrams for the cascade decays of LQs. The labels $P_1,\ldots,P_7$ denote the possible particles in the chain, and $\overline{P}_1$ is the anti-particle of $P_1$. All particles are specified in Tab.~\ref{tab:lq_cascade}. 
}
\label{fig:LQ_cascade}
\end{figure}

In Fig.~\ref{fig:LQ_cascade} and Tab.~\ref{tab:lq_cascade}, we show the processes $pp \to e^\pm e^\pm jj W^\pm$ with $j$ denoting a quark or anti-quark at the parton level in the UV models for the SMEFT operators $\mathcal{O}_1^{(9)}$, $\mathcal{O}_2^{(9)}$ and $\mathcal{O}_3^{(9)}$~\footnote{The SSDL process can also be achieved in the UV model for $\mathcal{O}_4^{9)}$ with a different topology of the decay channels, similar to that in Ref.~\cite{Graesser:2022nkv}. On the other hand, it is difficult to generate the SSDL process in the model for $\mathcal{O}_{\bar d u LLD}^{(7)}$.}. Interestingly, the $W$ boson, which comes from the decay of heavy particles due to the covariant derivative interaction, is unique in the two-step UV completions we consider.

\begin{table}[ht]
\tabcolsep=4pt
\renewcommand\arraystretch{1.5}
\caption{The decays of LQs in UV models for the SMEFT operators $\mathcal{O}_1^{(9)}$, $\mathcal{O}_2^{(9)}$ and $\mathcal{O}_3^{(9)}$ in Eq.~\eqref{eq:smeft-dim9}.
}
\begin{center}
\begin{tabular}{c|c|c|c|c|c|c|c}
\hline 
\hline
operator & $P_1$ & $P_2$ & $P_3$ & $P_4$ & $P_5$ & $P_6$ & $P_7$\\ \hline
\multirow{2}{*}{$\mathcal{O}_1^{(9)}$}      
& $\tilde{R}_2^{-1/3}$ & $W^{-}$ & $U_1^{2/3}$ & $e^{+}$ & $d$ & $e^+$ & $\bar{u}$ \\ 
& $U_1^{-2/3}$ & $W^-$ & $\tilde{R}_2^{+1/3}$ & $e^+$ & $\bar{u}$ & $e^+$ & $d$ \\ \hline
\multirow{2}{*}{$\mathcal{O}_2^{(9)}$}      
&$\bar{S}_1^{-2/3}$ & $W^-$ & $\tilde{V}_2^{1/3}$ & $e^+$ & $\bar{u}$ & $e^+$ & $d$ \\ 
&$\tilde{V}_2^{-1/3}$ & $W^-$ & $\bar{S}_1^{2/3}$ & $e^+$ & $d$ & $e^+$ & $\bar{u}$ \\ \hline
\multirow{1}{*}{$\mathcal{O}_3^{(9)}$} &$\tilde{R}_2^{-2/3}$ & $\bar{u}$ & $\Psi^0$ & $e^{+}$ & $W^-$ & $e^+$ & $d$  \\
\hline
\hline 
\end{tabular}
\end{center}
\label{tab:lq_cascade}
\end{table}

The pair production of LQs at the LHC is dominated by the gluon fusion $gg \to \mathrm{LQ}+\overline{\mathrm{LQ}}$. The cross section of $pp \to \mathrm{LQ}+\overline{\mathrm{LQ}}$ is expressed as
\begin{align}
\label{eq:xsec}
{\sigma}_{\mathrm{LQ}} = K \int \dfrac{d \hat{s}}{s} \int_\tau^1 \dfrac{dx}{x}  f_{g/p}(x) f_{g/p}\left(\dfrac{\tau}{x}\right)\hat{\sigma}_{\mathrm{LQ}}\;,
\end{align}
where $\sqrt{\hat{s}}$ is the center-of-mass (c.m.) energy of the parton subprocess,  $\sqrt{s}$ is the c.m. colliding energy, $f_{g/p}$ denotes the parton distribution function for the gluon in a proton, and $\tau \equiv \hat{s}/s$.
The leading-order cross sections at parton level $\hat{\sigma}_{\mathrm{LQ}}$ is~\cite{Kramer:1997hh,Workman:2022ynf}
\begin{align}
\label{crosssection}
\hat{\sigma}_{\mathrm{LQ}}=&\frac{\alpha_s^2 \pi}{96 \hat{s}} \left[\beta\left(41-31 \beta^2\right)\right.\nn\\
&\left.+\left(18 \beta^2-\beta^4-17\right) \ln \frac{1+\beta}{1-\beta}\right]\;.
\end{align}
Here,  $\beta \equiv$ $\sqrt{1-4 m_{\mathrm{LQ}}^2 / \hat{s}}$. In Eq.~\eqref{eq:xsec}, the $K$-factor is included to parameterize the next-to-leading order QCD corrections~\cite{Kramer:2004df,Mandal:2015lca,Dorsner:2016wpm}.

\begin{figure}[!htb]
\centering
\includegraphics[width=0.22\textwidth]{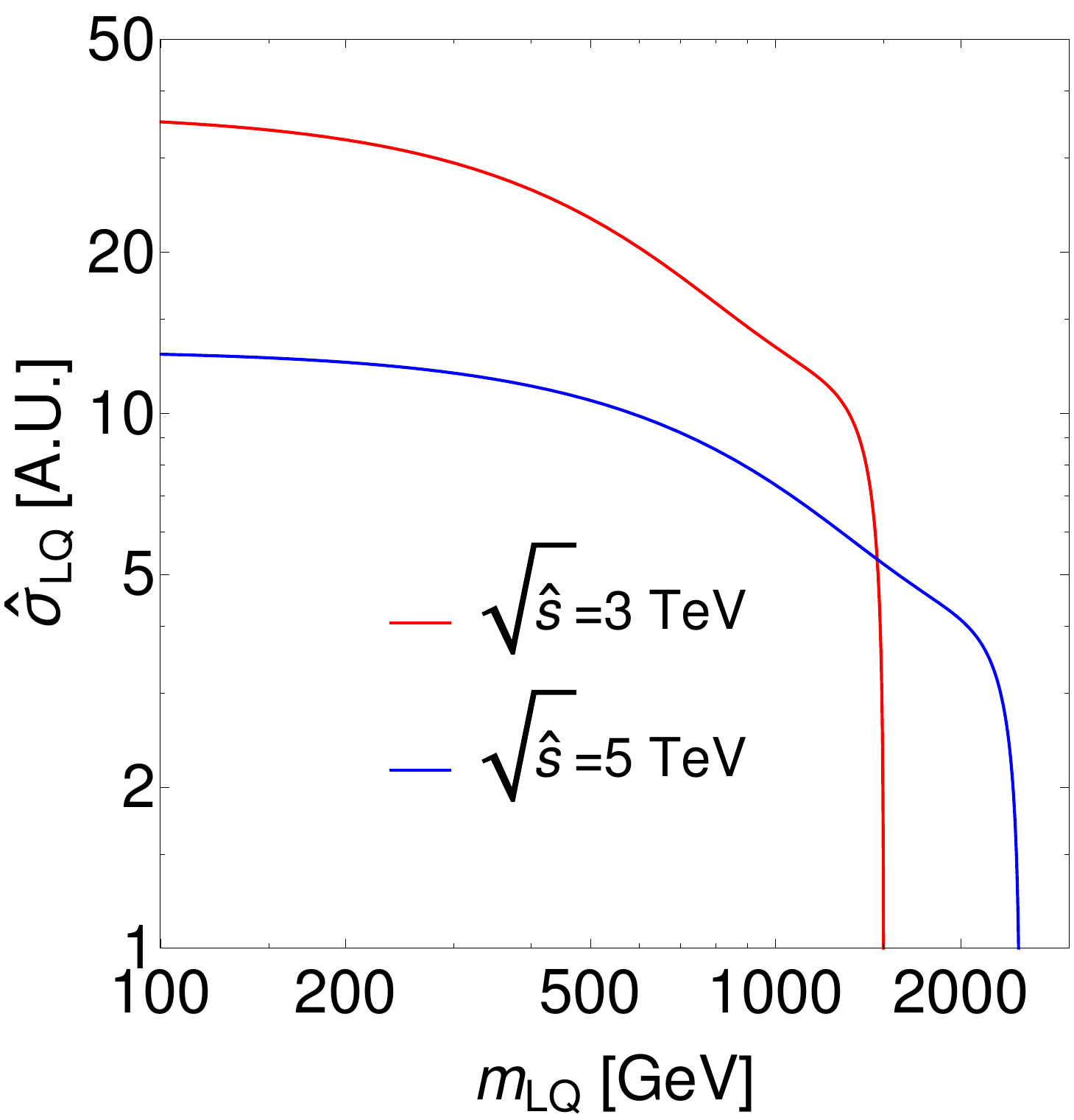}
\includegraphics[width=0.25\textwidth]{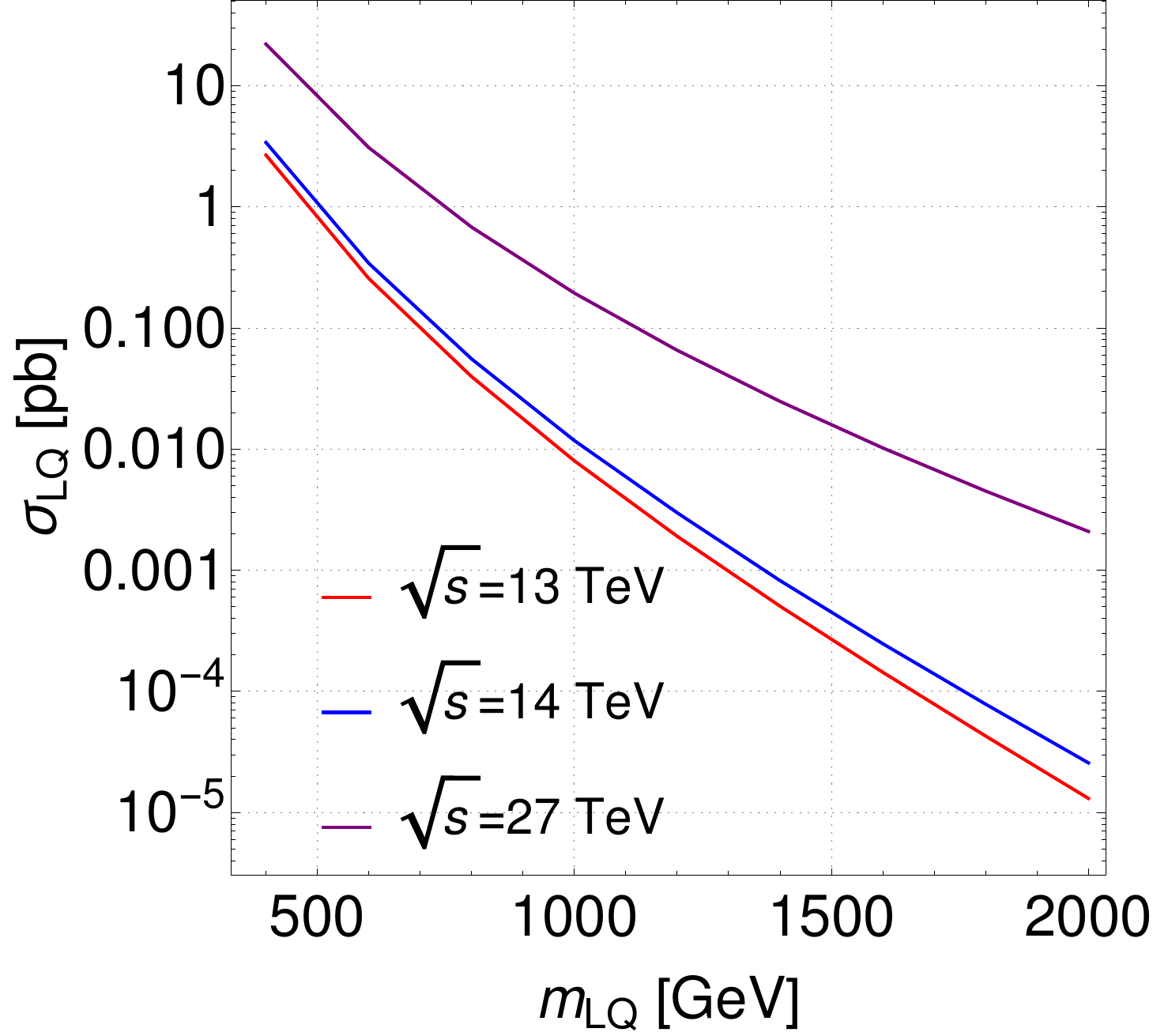}
\caption{Left: $\hat{\sigma}_{\rm LQ}$ in arbitrary unit (A.U.) with $\sqrt{\hat s} = 3$, $5\tev$ for different masses of LQ. Right: $\sigma_{\rm LQ}$ with $\sqrt{s} = 13$, 14, and $27\tev$ for different masses of LQ.
}
\label{fig:lq_prod_xsec}
\end{figure}

The partonic cross section $\hat{\sigma}_{\rm LQ}$ and hadronic cross section  $\sigma_{\rm LQ}$ are illustrated  in Fig.~\ref{fig:lq_prod_xsec}. In the left panel, two benchmark values of the partonic energy $\sqrt{\hat s}$ are chosen.  In both scenarios, $\hat{\sigma}_{\rm LQ}$ drops rapidly with the increase of LQ mass dubbed $m_{\rm LQ}$. Besides, for $m_{\rm LQ}\gtrsim 1.8\tev$, the increase of $\sqrt{\hat{s}}$ from $3\tev$ to $5\tev$ can significantly enhance the gluon-fusion subprocess of LQ pair production. In the right panel, we
can see that $\sigma_{\rm LQ} \simeq 1.5 \times 10^{-2}\fb$ at $\sqrt{s}=13\tev$, while $\sigma_{\rm LQ}$ reaches about $2\fb$ at $\sqrt{s}=27\tev$. From Fig.~\ref{fig:lq_prod_xsec}, we find that in order to search for LQ with its mass around $2\tev$ or heavier, it would be critical and effective to improve the colliding energy.

There have been extensive searches for the SSDL signature at the 13~TeV LHC in the benchmark models of supersymmetric particles~\cite{ATLAS:2019fag, CMS:2020cpy} or heavy Majorana neutrinos~\cite{CMS:2018agk,CMS:2018jxx,CMS:2021dzb,ATLAS:2018dcj,ATLAS:2023cjo} with null results, which can be re-interpreted as searches for the UV resonances of the $\onbb$-decay related operators.
Besides, we will consider SSDL searches at the high-luminosity LHC (HL-LHC) with $\sqrt{s}=14\tev$ and the proposed high-energy upgrade (HE-LHC) with $\sqrt{s}=27\tev$. 

As a case study, we will consider the process $pp \to \tilde R_2^{1/3} \tilde R_2^{-1/3}$ and the decay chains in the first row in Tab.~\ref{tab:lq_cascade} where the leptoquark $U_1$ and $W$ boson are on the shell. The partial decay widths are
\begin{align}
\label{RVW}
&\Gamma(\tilde R_2^{-1/3}\rightarrow U_1^{2/3}+ W^-) =\frac{\beta(m_R^2, m_U^2, m_W^2) }{64\pi m_R^3m_U^2} \lambda_{DH}^2 \\
&\qquad \times \left[(m_R^2-m_U^2-m_W^2)^2+8m_U^2m_W^2 \right]\;,
\end{align}
where $\beta(x, y, z)\equiv \left[(x-y-z)^2-4 y z\right]^{1/2}$, and
\begin{align}
\label{Rue}
\Gamma(\tilde R_2^{-1/3}\rightarrow u+e^-)=\frac{\lambda_{u\Psi}^2 \sin^2{\theta}}{16\pi}m_R\;.
\end{align}	
The mixing angle $\theta$ is defined as
\begin{align}
\label{eq:sth-O1}
\sin\theta = \dfrac{f_{\Psi e}v}{\sqrt{2} m_\Psi}\;,
\end{align}
which is convenient for the phenomenological study, see Appendix~\ref{app:mixing} for more discussions. In the model for $\mathcal{O}_1^{(9)}$, $U_1^{2/3}$ can only decay into $e^+$ and $d$. Thus the cross section of the signal $pp\to e^\pm e^\pm j j W^\mp$  at the parton level is expressed as 
\begin{align}
\label{eq:xsec-signal}
\sigma_s = 2\ \sigma_{\rm LQ}\times \mathcal{B}_1 \times \mathcal{B}_2\;,
\end{align}
where $j$ denotes a quark or anti-quark,  $\mathcal{B}_{1}$ and $\mathcal{B}_2$ are the branching rations of $\tilde R^{-1/3}\rightarrow U_1^{2/3}+ W^-$ and $\tilde R^{1/3}\rightarrow \bar u+e^+$, respectively, and the $W$ boson can decay hadronically or leptonically.

As an estimate, taking $m_R = 2\tev$, $m_U =1.8\tev$, and we obtain
\begin{align}
\dfrac{\mathcal{B}_2}{ \mathcal{B}_1} \simeq \left( \dfrac{\sin\theta}{0.05}\dfrac{\lambda_{u\Psi}}{\lambda_{DH}} \right)^2\;.
\end{align}
If $\sin\theta = 0.05$, the coupling and mass of $\tilde R_2$ is $f_{\Psi e}/m_\Psi \simeq 1/(3.5\tev)$.

The main SM backgrounds of SSDL searches include prompt backgrounds $WW$, $WZ$ and $ZZ$, jet fake backgrounds from $j\to e$, and charge flip backgrounds from the misidentification of electron charge. 
We find that the selection criteria in the search for heavy Majorana neutrino at the LHC Run 2 with the integrated luminosity of $139\fbi$ in the resolved channel~\cite{ATLAS:2023cjo} are suitable for our signal. 
The signal events are generated 
using \texttt{MadGraph5\_aMC@NLO}~\cite{Alwall:2014hca}, 
which are passed to
\texttt{Pythia8}~\cite{Sjostrand:2014zea} and \texttt{Delphes3}~\cite{deFavereau:2013fsa} for parton shower and detector simulation, respectively. 
A pair of same-sign electrons and at least two jets are selected if~\cite{ATLAS:2023cjo}
\begin{align}
\label{eq:cut1}
&p_T^{e1(2)} > 40~(25)\gev\;, && \mid \eta_e \mid < 2.47\;,\nn\\
&p_T^j > 100\gev\;, && \mid \eta_j \mid < 2.5\;,
\end{align}
where $p_T^{e1}$, $p_T^{e2}$ and $p_T^j$ are the transverse momenta of the leading and sub-leading electrons and jets, respectively, and $\eta_e$ and $\eta_j$ are their pseudo-rapidities.

\begin{figure}[!htb] 
\centering
\includegraphics[width=0.23\textwidth]{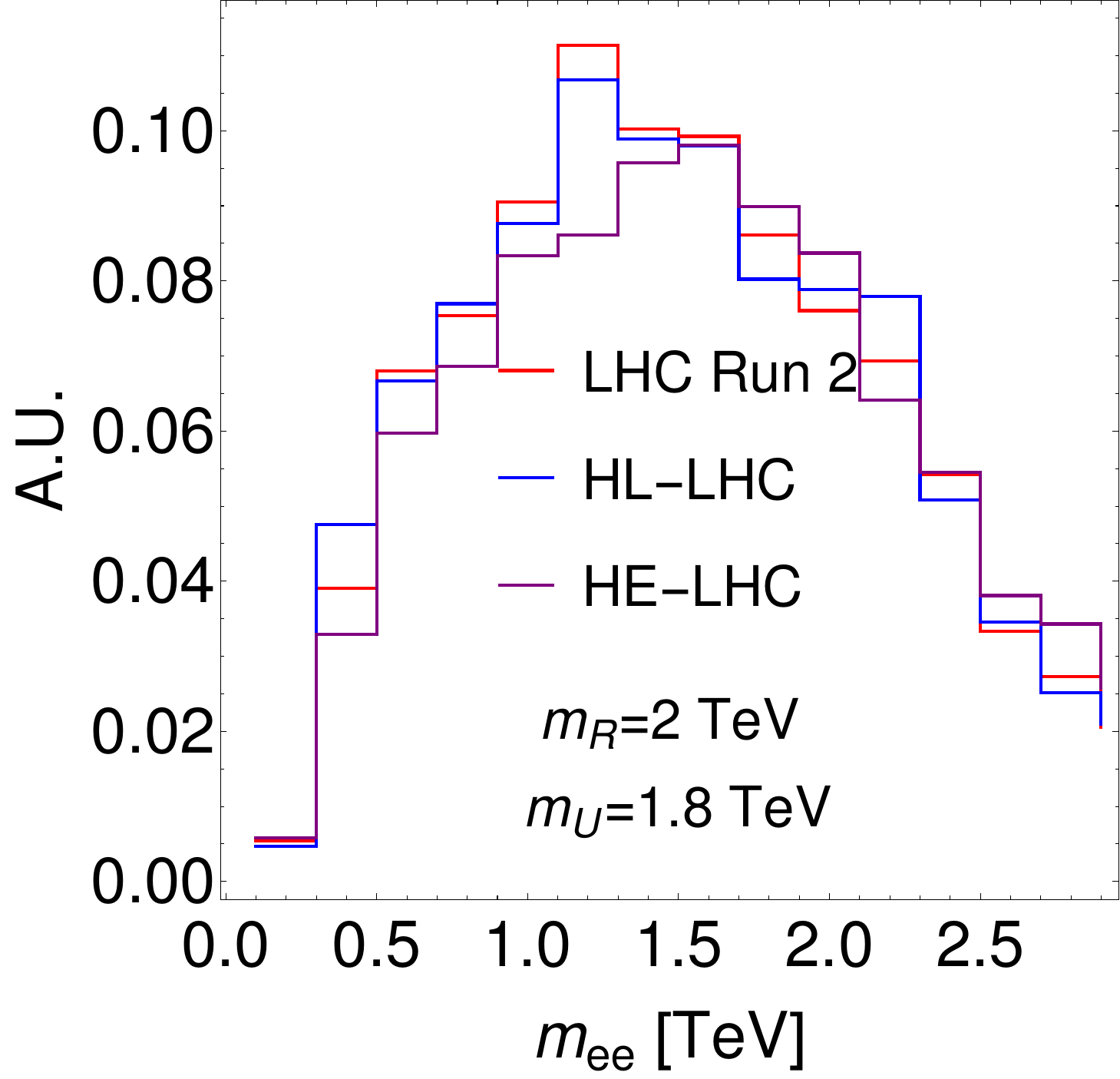}	
\includegraphics[width=0.23\textwidth]{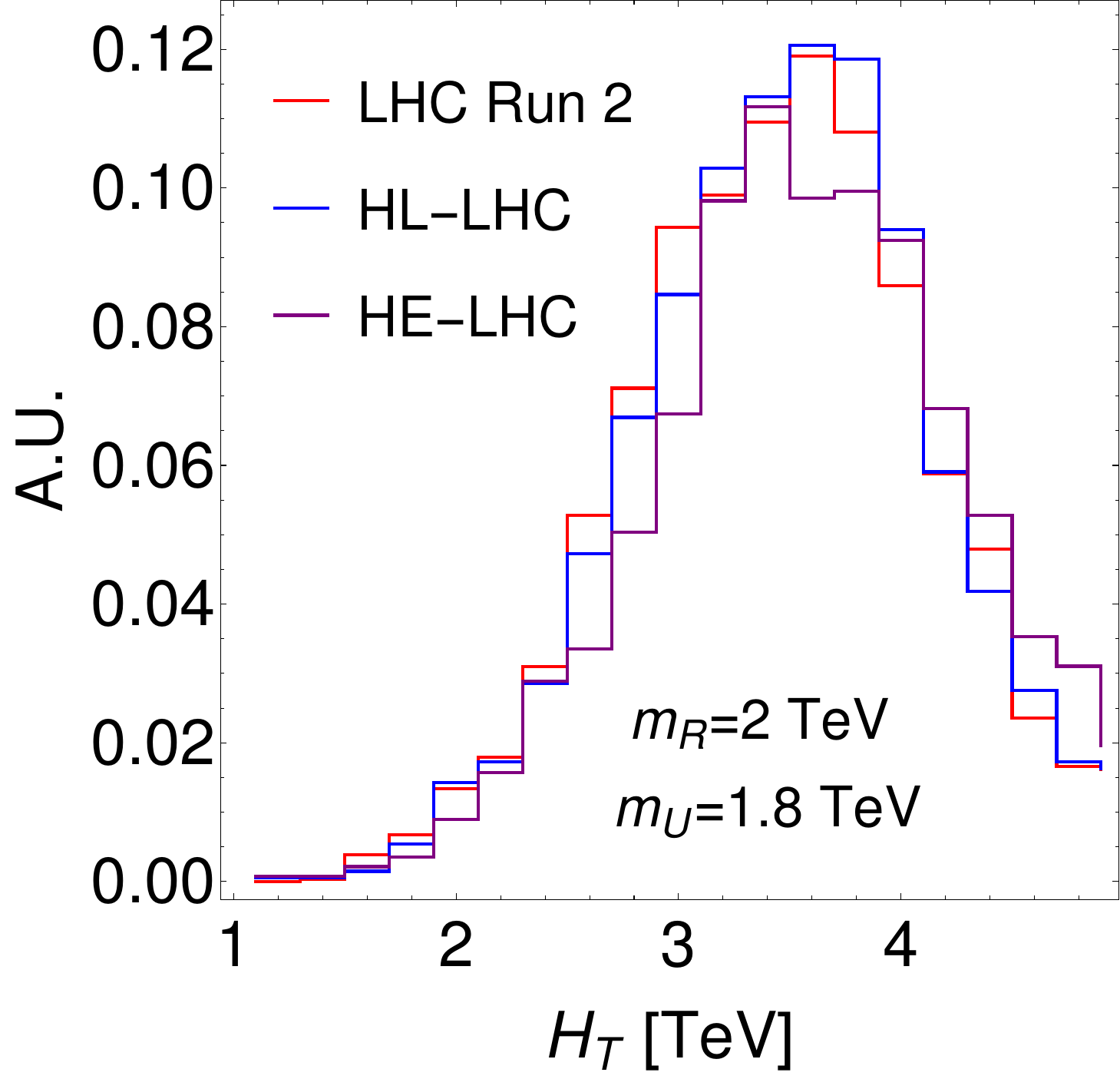}	
\caption{Kinematic distributions of the signals at the LHC Run 2, HL-LHC and HE-LHC with $\sqrt{s}=13$, 14 and $27\tev$ respectively after passing the selection cuts in Eq.~\eqref{eq:cut1}.
}
\label{fig:kin-LHC}
\end{figure}

The $m_{ee}$  and $H_T$ distributions of the signal after the cuts in Eq.~\eqref{eq:cut1} are displayed in Fig.~\ref{fig:kin-LHC}, where $m_{R}=2$ TeV, $m_U=1.8$ TeV are assumed. We can see that the signal has large $m_{ee}$ and $H_T$, which is reasonable due to heavy resonances $\tilde R_2$ and $U_1$. 

In the ATLAS analysis~\cite{ATLAS:2023cjo}, the cuts invariant mass of electron pair $m_{ee}>400\gev$ and the scalar sum of the transverse momenta of electrons and two most energetic jets $H_T > 400\gev$ are further imposed to reduce the SM backgrounds. 
About 40 background events are left, which can be counted in the $H_T$ distribution in Fig.5(b) of Ref.~\cite{ATLAS:2023cjo}, which also indicates that less than 1 SM background event for $H_T > 1.6\tev$.

In order to reject most of the SM backgrounds, we require a harder cut $H_T > 3\tev$
for the searches at the 13~TeV LHC Run 2, 14~TeV HL-LHC and 27~TeV HE-LHC. The corresponding signal selection efficiencies after passing the selection cuts and this optimized cut are $\epsilon_s = 0.29$, 0.30 and $0.32$, respectively. We emphasize that a more delicate analysis with a stronger cut on $m_{ee}$ can also be used to effectively remove the small SM backgrounds with less impact on the signals. 

The number of signal events after passing all cuts is
\begin{align}
n_s = \sigma_{s} \epsilon_s \mathcal{L} \;.
\end{align}
where $\sigma_s$ is the signal cross section obtained in Eq.~\eqref{eq:xsec-signal} and $\mathcal{L}$ denotes the integrated luminosity.  
We obtain that
$n_s$ is smaller than 1 at the LHC Run 2 implying that current SSDL searches at still weak to constrain the LNV parameters for the UV model we discuss.

The 95\% C.L. exclusion limit in case of no SM background is evaluated by requiring that the number of signal events $n_s = 3$~\cite{Junk:1999kv,Bhattiprolu:2020mwi}. In the next section, we will consider the exclusion limits that could be obtained at the HL-LHC and HE-LHC.

\section{Results and discussions}
\label{sec:result}

In this section, we study the complementary searches for the UV resonances in the models in $\onbb$ decay and LHC. For illustration, the sensitivities on the model for $\mathcal{O}_1^{(9)}$ will be compared.
From Eqs.~\eqref{eq:matching1}~\eqref{eq:sth-O1}, the square root of the inverse half-life
\begin{align}
\label{eq:half-life-estimate}
\left( T_{1/2}^{0\nu}\right)^{-1/2}&\propto  \dfrac{\lambda_{ed} \lambda_{DH}\lambda_{u\Psi}  \sin\theta }{m_U^2 m_R^2} \;,
\end{align}

The signal cross section depends on the masses via the LQ pair production and the decay branching ratios. Taking $m_R = 2\tev$ and $m_U = 1.8\tev$, we have that square root of the signal cross section
\begin{align}
\label{eq:cross-section-estimate}
\sigma_s^{1/2}&\propto \dfrac{\lambda_{DH}\lambda_{u\Psi}  \sin\theta}{\left(\sin\theta \lambda_{u\Psi}\right)^2 + \left( 0.05 \lambda_{DH} \right)^2 } \;.
\end{align}

From Eqs.~\eqref{eq:half-life-estimate}~\eqref{eq:cross-section-estimate}, we can see that the signal process at the LHC is insensitive to the parameter $\lambda_{ed}$, while both the signal cross section and the $\onbb$-decay rate are suppressed by the mixing angle $\sin\theta$. The dependence on $\lambda_{u\Psi}$ and $\lambda_{DH}$ implies that $\onbb$-decay and LHC searches have different sensitivities to the parameters.

\begin{figure}[h]
\centering
\includegraphics[width=0.45\linewidth]{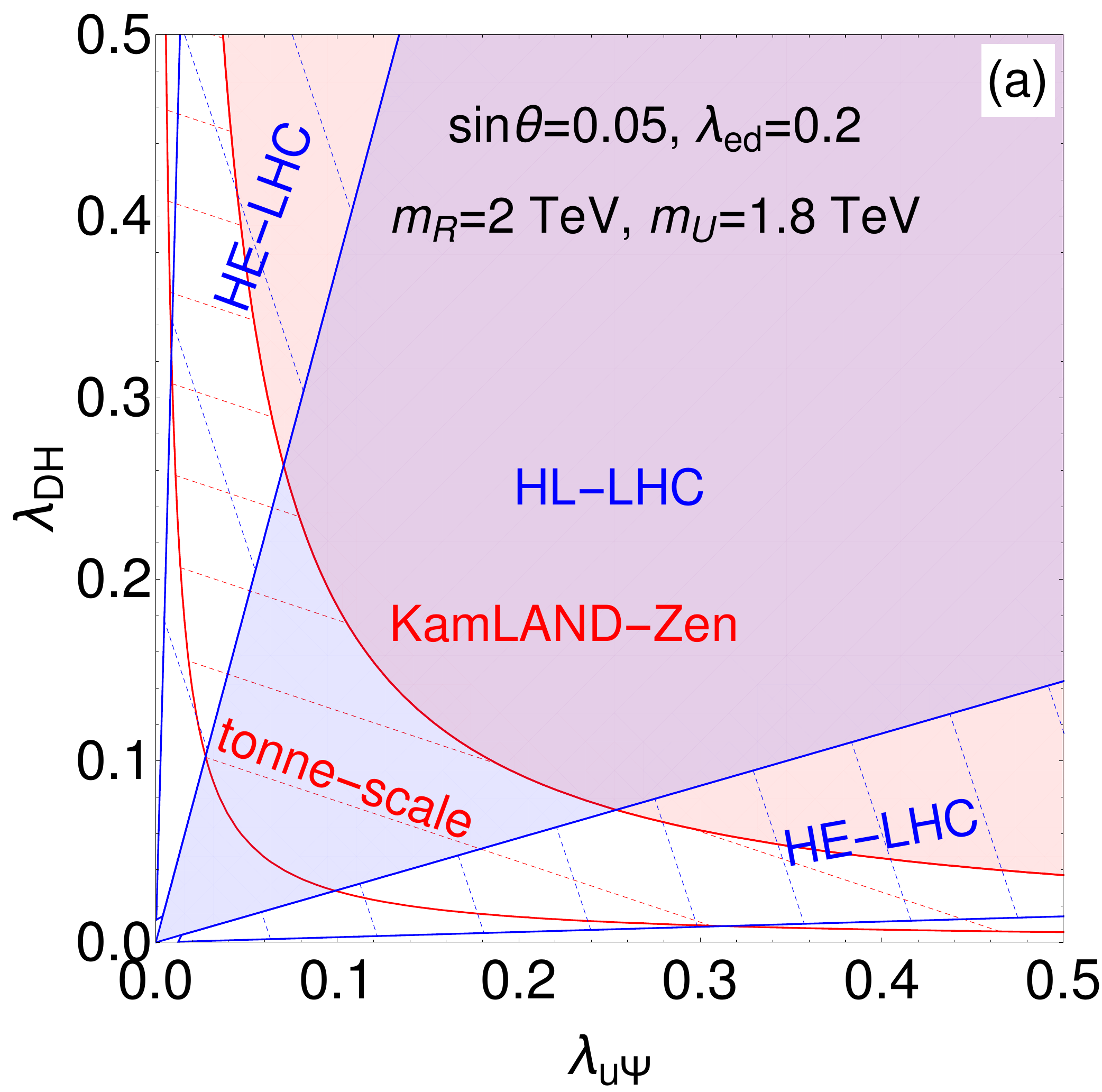}
\includegraphics[width=0.45\linewidth]{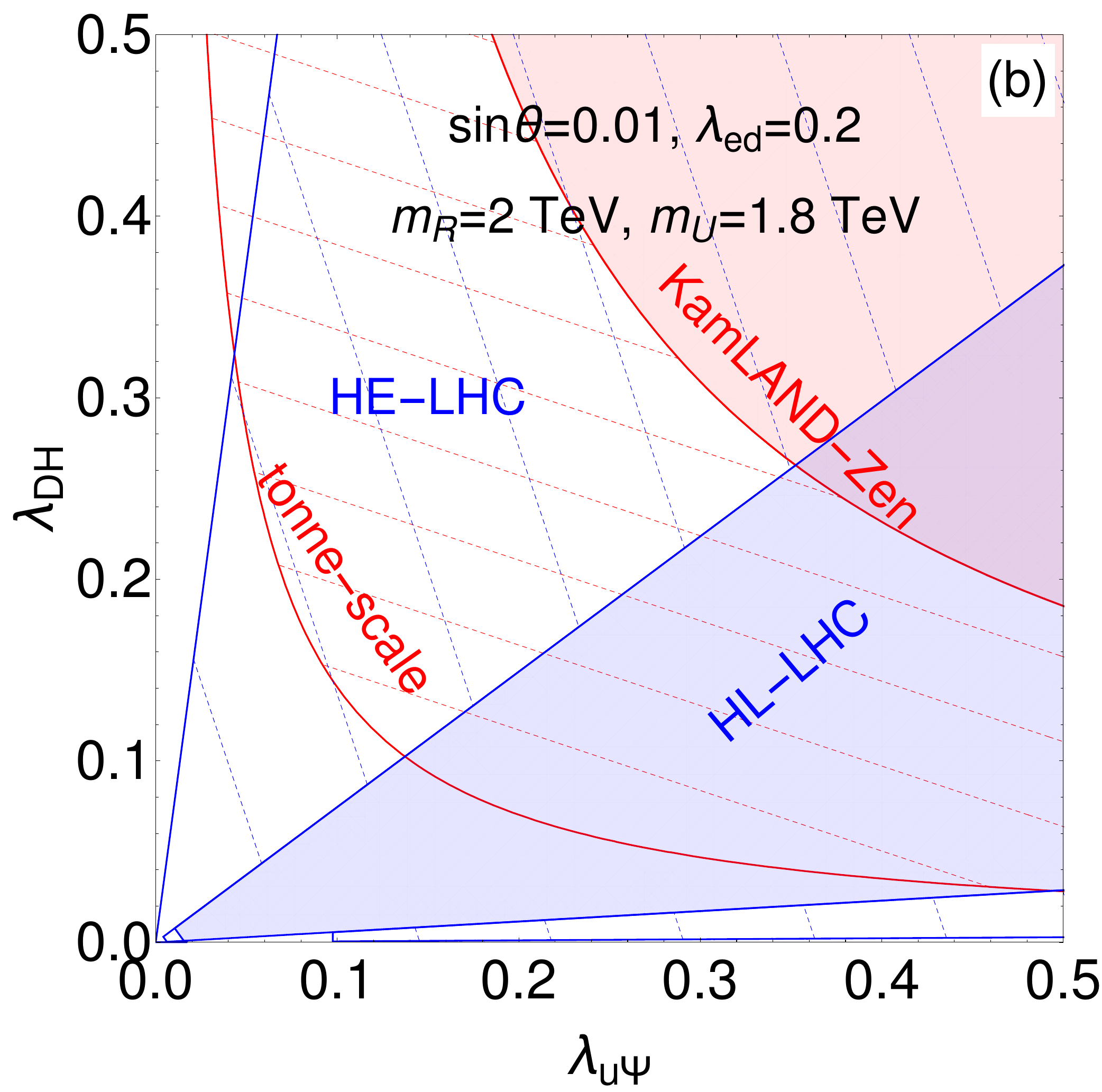}
\includegraphics[width=0.45\linewidth]{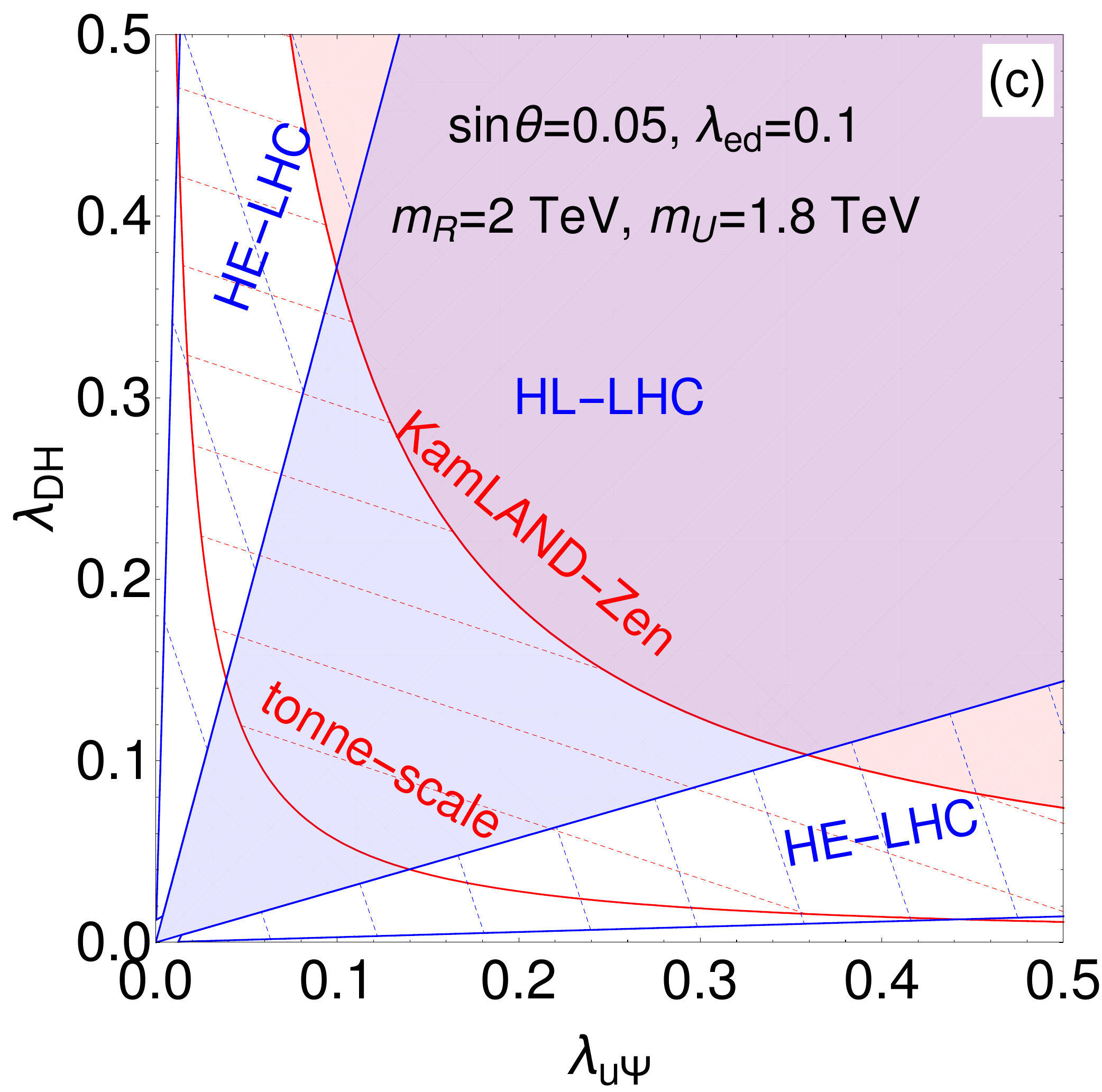}
\includegraphics[width=0.45\linewidth]{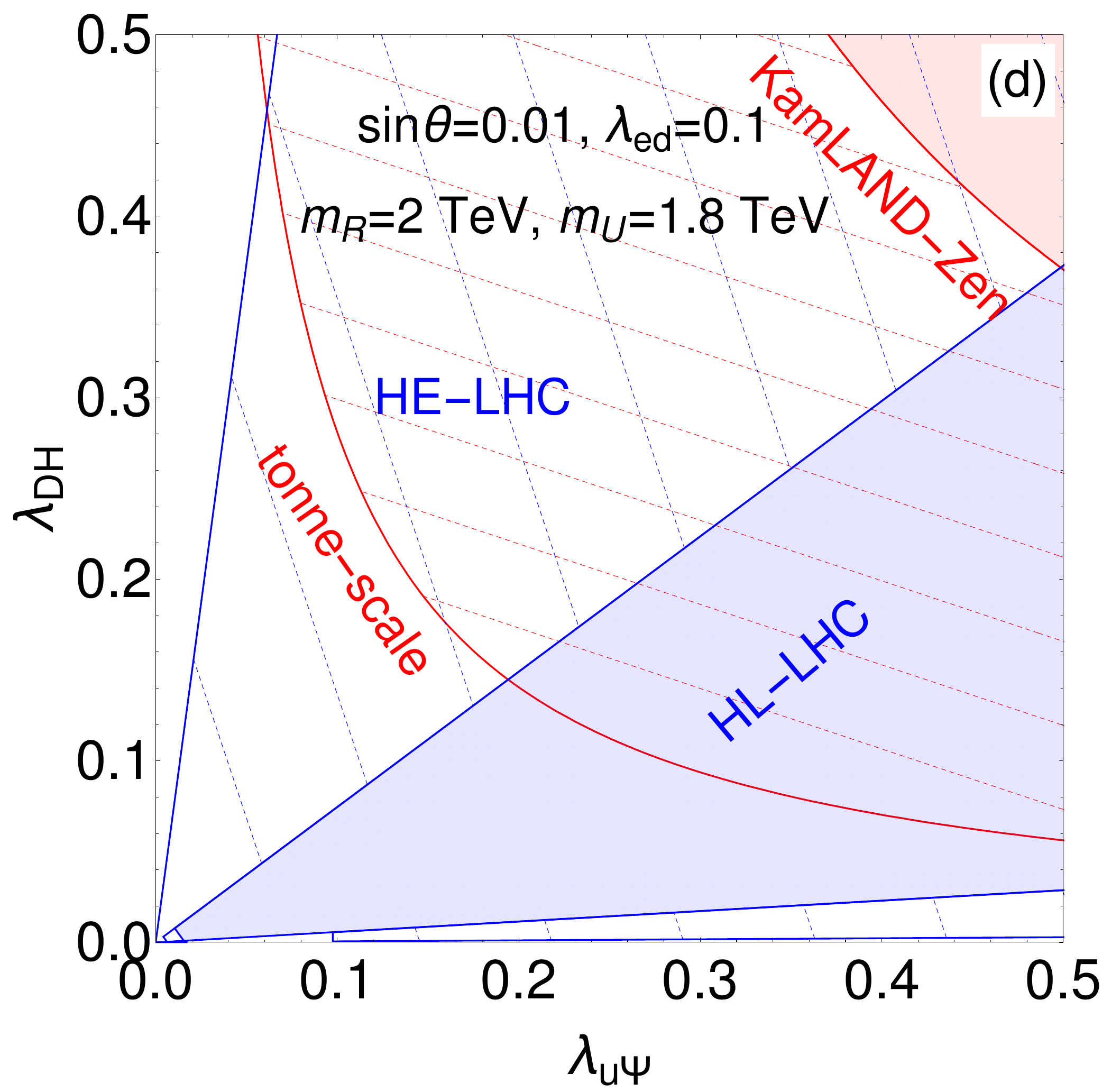}
\caption{The sensitivities in the plane of $\lambda_{u\Psi}$ and $\lambda_{DH}$ of KamLAND-Zen (red) and tonne-scale (red slash shading) $\onbb$-decay and SSDL searches at the HL-LHC (blue) and HE-LHC (blue slash shading). See text for more details.}
\label{fig:combined}
\end{figure}

In Fig.~\ref{fig:combined}, we show the combined sensitivities to the couplings $\lambda_{u\Psi}$ and $\lambda_{DH}$
in the KamLAND-Zen and future tonne-scale $\onbb$-decay experiments as well as at the HL-LHC and HE-LHC. 
The red regions are excluded by the $\onbb$-decay search in the KamLAND-Zen experiment at 90\% C.L., while the blue regions are expected to be excluded by the SSDL searches at the HL-LHC with the integrated luminosity of $\mathcal{L} = 3\abi$ at 95\% C.L.. The slash shading regions in red and blue denote those that can be further excluded in future tonne-scale $\onbb$-decay experiments and HE-LHC with $\mathcal{L} = 3\abi$, respectively.

We fix the masses of LQs as $m_R = 2\tev$ and $m_U = 1.8\tev$ and assume that $m_\Psi > m_R$. Four benchmark scenarios that satisfy the existing constraints~\cite{Crivellin:2021egp,ATLAS:2020dsk}
are considered: (a) $\sin\theta = 0.05$, $\lambda_{ed}=0.2$; (b) $\sin\theta = 0.01$, $\lambda_{ed}=0.2$; (c) $\sin\theta = 0.05$, $\lambda_{ed}=0.1$; (d) $\sin\theta = 0.01$, $\lambda_{ed}=0.1$.

In all of the scenarios, searches for the UV resonances -- the LQs $\tilde R_2$ and $U_1$ -- in the $\onbb$ decay and at the LHC  are complementary to each other. Besides, we can see that the sensitivities of the high-energy upgrade of the LHC, i.e., HE-LHC, are much improved compared to the HL-LHC, because the cross section of LQ production increases significantly at the HE-LHC. 

From panels (a) and (c), most of the parameter space is in the reach of HE-LHC and tonne-scale $\onbb$-decay experiments if $\sin\theta = 0.05$, or equivalently $f_{\Psi e}/m_\Psi \simeq 1/(3.5\tev)$. For a larger $m_\Psi$ or smaller $f_{\Psi e}$, both the sensitivities of LHC and $\onbb$-decay experiments are reduced. In this case, the HE-LHC and tonne-scale $\onbb$ experiments are crucial to probe the couplings of the LQs, as seen in panels (b) and (d). In the comparison of panels (a) and (c) as well as (b) and (d), the reaches of $\onbb$-decay searches are sensitive to the coupling $\lambda_{ed}$, thus it can be constrained alongside with $\lambda_{u\Psi}$ and $\lambda_{DH}$.

Finally, we comment that if the LQs have larger masses, the sensitivities of $\onbb$-decay searches are less impacted compared to the LHC searches, because the cross section of LQ pair production drops rapidly with the increase of the LQ mass, as clearly shown in Fig.~\ref{fig:lq_prod_xsec}. The interplay for other choices of $m_R$ and $m_U$, or in the UV models for $\mathcal{O}_2^{(9)}$ and  $\mathcal{O}_3^{(9)}$ can be studied analogously.

\section{Conclusion}
\label{sec:conclusion}

In this work, we have investigated the two-step UV completions of the effective operators that give rise to chirally enhanced contributions to $\onbb$ decay. There are one dim-7 and four dim-9 SMEFT operators that can be matched to 
the $\Delta L =2$ quark-lepton $O_{4X}^{(9)}$ with $X=L$ or $R$. 
We have introduced possible UV completions for each of the relevant SMEFT operators with the leptoquarks (LQs), and studied the searches for the UV resonances at the LHC.

In order to illustrate the complementarities of $\onbb$-decay and LHC searches, we study in detail the UV model for $\mathcal{O}_1^{(9)}$. Assuming the new physics couplings $g_{\rm NP} =0.2$ or 0.5 (weakly-coupled), the $\onbb$-decay experiments are sensitive to the LNV scale $\Lambda \sim 2-3\tev$ or $4-5\tev$, respectively. On the other hand, since the LQ
production cross section is reduced significantly with the increase of the LQ mass, LHC Run 2 is unable to constrain the related LNV parameters. 
We thus consider the same-sign dilepton searches for the process $pp\to e^\pm e^\pm j j W^\mp$ at the high-luminosity LHC (HL-LHC) and high-energy LHC (HE-LHC) with the integrated luminosities of $3\abi$.

We obtain that the direct searches at the HL-LHC and HE-LHC and indirect searches in the KamLAND-Zen and future tonne-scale $\onbb$-decay experiments are complementary to each other in testing the UV completions of the relevant SMEFT operators. Thus possible $\onbb$-decay signals from chirally enhanced mechanisms
can be diagnosed with the LHC searches.

\begin{acknowledgments}
We would like to thank Xiao-Dong Ma and Zhe Ren for very helpful discussions.
This work is supported by 
the National Natural Science Foundation of China under Grant No. 12347105.
GL and XZ are partly supported by Fundamental Research Funds for the Central Universities in Sun Yat-sen University (23qnpy62), and SYSU startup funding.
J.H.Y. is supported in part by the National Science Foundation of China under Grants No. 12022514, No. 12375099 and No. 12047503, and National Key Research and Development Program of China Grant No. 2020YFC2201501, and No. 2021YFA0718304. 
\end{acknowledgments}

\appendix

\section{Mixing of lepton fields}
\label{app:mixing}

In this appendix, the mixing of the SM lepton and vector-like fermion fields will be discussed.
For illustration, we consider the UV model for $\mathcal{O}_1^{(9)}$.
The mass terms of vector-like fermions and the mixing with the SM lepton fields via the Yukawa interactions are given by 
\begin{align}
   \mathcal{L} \supset y_{e} \bar{L} H e_R+f_{\Psi e} \bar{\Psi}_L H e_R+m_{\Psi} \bar{\Psi}_L \Psi_R+\text {h.c.}\;, 
\end{align}
where a mass term $\bar{\Psi}_R L$ can be rotated away~\cite{Kearney:2012zi,Graesser:2016bpz}. 
 
After the Higgs field develops a vacuum expectation value, we can obtain the mass terms of the charged leptons as follows
\begin{align}
\label{mix}
\mathcal{L}_{m_e }&=\begin{pmatrix}
			\bar{e}_L,{\bar{E}_L}
		\end{pmatrix}\mathcal{M}_e
		\begin{pmatrix}
			{e_R}\\{{E_R}}
		\end{pmatrix} \;,
\end{align}
where the mass matrix is given by
\begin{align}
\mathcal{M}_e&=\frac{1}{\sqrt{2}}
		\begin{pmatrix}
			{y_ev}&\ 0\\
			{f_{\Psi e}v}&\ \sqrt{2}m_{\Psi}
		\end{pmatrix}\;,
\end{align}
The mass matrix $\mathcal{M}_e$ can be diagonalized to yield mass eigenstates of charged fermions labeled by $e_R^{\prime}, E_R^{\prime}$, which are expressed as
\begin{align}
\label{mixs}
e_R^\prime &= \cos\theta e_R -\sin\theta E_R\;,\nonumber\\
E_R^\prime &= \sin\theta e_R +\cos\theta E_R\;,
\end{align}
where we define $\sin\theta = {f_{\Psi e}v}/(\sqrt{2} m_\Psi)$. 

The interactions between the vector-like fermions and the Higgs boson could modify the the Higgs couplings~\cite{Joglekar:2012vc,Kearney:2012zi}, while the constraints depend on other possible heavy particles in the UV theories, the detailed study of which is beyond the scope of this work.

\section{SMEFT operators involving \tf{$D_{\mu}L$}{DL}}
\label{app:CDE}

As mentioned in Sec.~\ref{sec:UV-completion}, the covariant derivatives of the SM fields are tracked from the kinematic terms of $\Psi$. 
In the following, we will explain how to obtain the SMEFT operators in Eq.~\eqref{eq:CD3}. 

From the interactions in Eq.~\eqref{eq:lag3}, we can solve the classical equation of motions (EOMs) for the heavy fields. Since there are several fields, one can first integrate one, and then the others. By using the EOMs we have
\begin{align}
\label{EOM}
\Psi_R&= -\frac{1}{m_{\Psi}^2}i\slashed{D}\left (\lambda_{u\Psi}u_R^c\tilde{R}_2+f_{L\Psi}L S^*\right)\nonumber\\
&= \frac{1}{m_{\Psi}^2}i\slashed{D}\left[-\frac{\lambda_{u\Psi}\lambda_{Ld}}{m_{R}^2}u_R^c(\epsilon \bar{L}d_R) \right.\nn\\
&\left.\qquad \qquad +\frac{f_{L\Psi}\mu}{m_S^2}L(H^{\dagger}H)\right]\;.
\end{align}
After combining the two terms on the right-hand side, we can obtain the effective interactions in Eq.~\eqref{eq:CD3}.

\begin{figure}[h]
	\centering
	\includegraphics[width=0.7\linewidth]{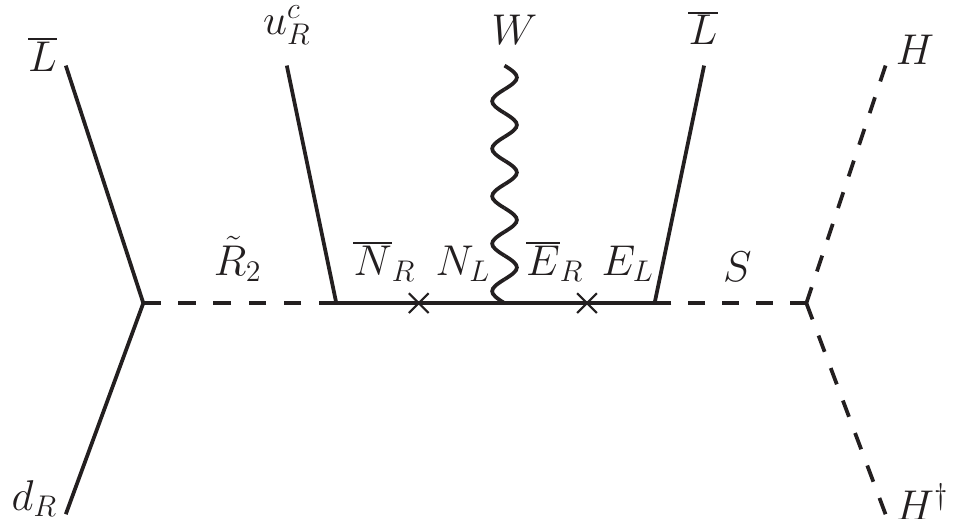}
	\caption{Feynman diagram for the UV completion of $\mathcal{O}_3^{(9)}$ with the vector-like fermions and $W$ boson being specified. }
	\label{fig:model3b}
\end{figure}

\begin{figure}[h]
	\centering
	\includegraphics[width=0.7\linewidth]{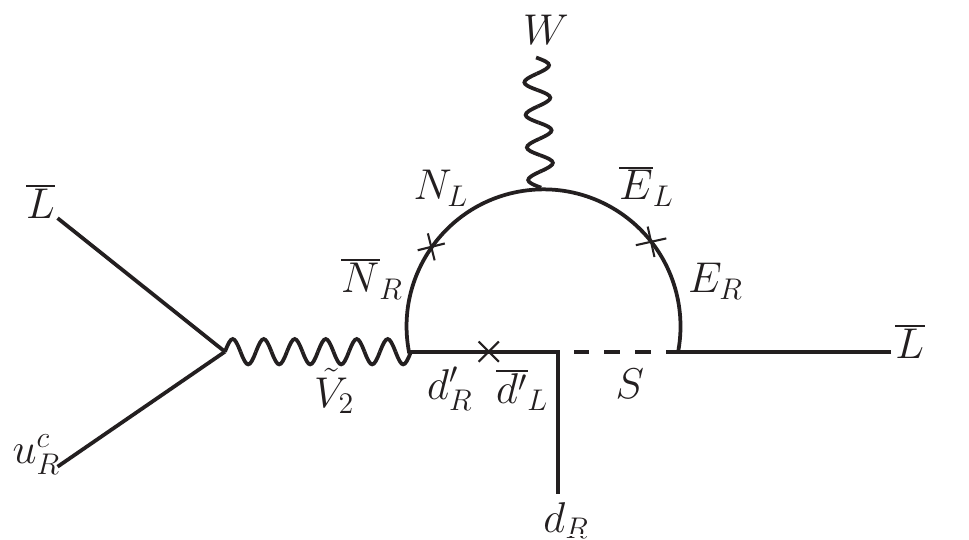}
	\caption{Feynman diagram for the UV completion of $\mathcal O_{\bar d u LLD}^{(7)}$ with the vector-like fermions, colored fermion and $W$ boson being specified.}
	\label{fig:model5b}
\end{figure}

In the diagrammatic approach, the SMEFT operator $\mathcal{O}_1^{(9)}$ can be generated by integrating out the heavy fields in Fig.~\ref{fig:model3b}, which is more specific than Fig.~\ref{fig:model2}. The components of the vector-like fermions $\Psi$ with mass insertion are explicitly shown, and the $W$ boson is attached. Similarly, the UV completion of the dim-7 SMEFT operator $\mathcal O_{\bar d u LLD}^{(7)}$ is specified in Fig.~\ref{fig:model5b}. 


\section{Neutrino masses }
\label{app:neutrino-mass}
 As mentioned above, the UV models for the relevant SMEFT operators
give negligible contributions to the neutrino masses. To see it, we draw the corresponding Feynman diagrams of the light neutrino Majorana masses, and estimate their contributions.

The leading contribution to Majorana masses in the UV model for $\mathcal{O}_{3}^{(9)}$ is generated at the two-loop level, while those for $\mathcal{O}_{1,2,4}^{(9)}$ and $\mathcal{O}_{\bar{d}uLLD}^{(7)}$ are generated at the three-loop level.

\begin{figure}[h]
 	\centering
 	\includegraphics[width=0.7\linewidth]{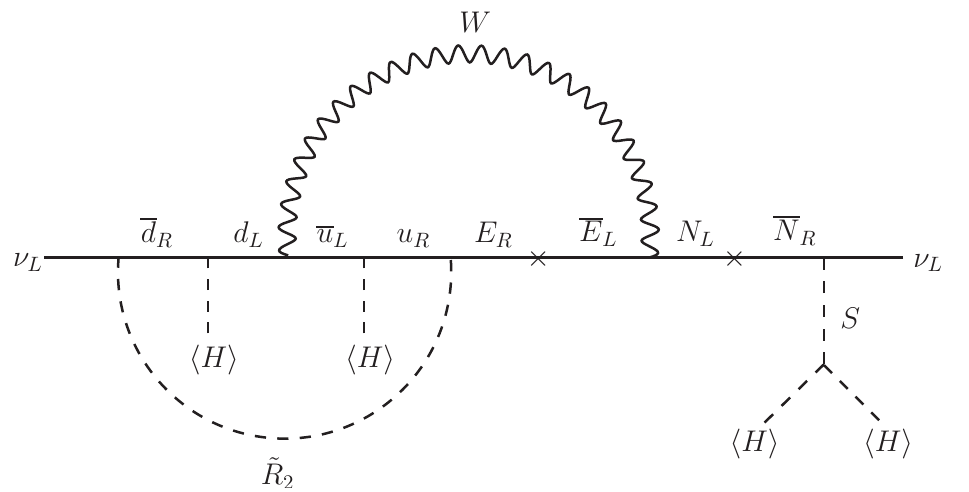}
 	\caption{Feynman diagram of the light neutrino Majorana masses in the UV model for $\mathcal{O}_3^{(9)}$. }
	\label{2loopmass}
 \end{figure}

In Fig.~\ref{2loopmass}, we show Feynman diagram of the light neutrino mass in the UV model for $\mathcal{O}_3^{(9)}$. The contribution is estimated as
\begin{align}
m_{\nu}&\sim \frac{m_u m_d  v^2 \mu}{(16\pi^2)^2}\frac{\lambda_{Ld}\lambda_{u\Psi}f_{L\Psi}}{m_{S}^2m_R^2m_{\Psi}^2}\Lambda_{\text{UV}}^2\nonumber\\
&\approx 2.8\times 10^{-17} \mu
\left(\frac{\Lambda_{\text{UV}}}{1 \text{TeV}}\right)^2\left(\frac{1 \text{TeV}}{\Lambda}\right)^6\;.
\end{align}
Here, $\Lambda_{\rm UV}$ is the UV cutoff $\Lambda_{\text{UV}}\sim \Lambda$, and we have set $\lambda_{Ld}\lambda_{u\Psi}f_{L\Psi}\equiv 1$, $\Lambda^6=m_{S}^2m_R^2m_{\Psi}^2$ in the second line.
We can see that $m_\nu$ is negligible even for $\mu \sim 1\tev$.

 \begin{figure}[h]
	\centering
	\includegraphics[width=0.7\linewidth]{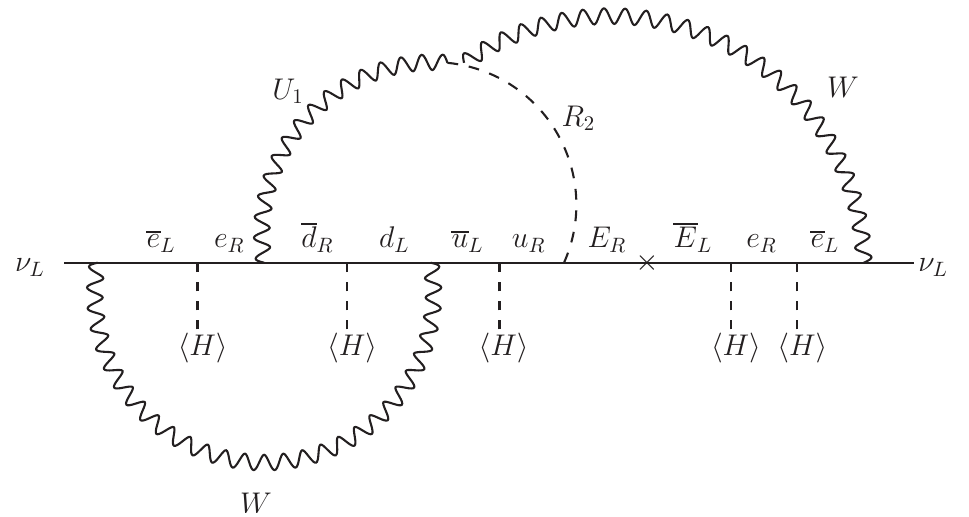}
	\caption{Feynman diagram of the light neutrino Majorana masses in the UV model for $\mathcal{O}_1^{(9)}$.}
	\label{3loopmass}
\end{figure}

In Fig.~\ref{3loopmass}, we show Feynman diagram of the light neutrino mass in the UV model for $\mathcal{O}_1^{(9)}$, and those for $\mathcal{O}_2^{(9)}$, $\mathcal{O}_4^{(9)}$ and $\mathcal{O}_{\bar{d}uLLD}^{(7)}$ can be obtained analogously.
In all of these cases, Majorana masses of light neutrinos generated at the three-loop level can be neglected.



\bibliographystyle{apsrev4-1}
\bibliography{reference}
\end{document}